\documentclass[pra,10pt,final,twocolumn,showpacs,groupedaddress,floatfix]{revtex4}
\usepackage[dvips]{epsfig}
\usepackage{float}
\usepackage{amsfonts}
\usepackage{amsmath}
\usepackage{amssymb}
\usepackage{enumerate}

\usepackage{hhline}
\usepackage{threeparttable}
\usepackage{supertabular}
\usepackage{multirow}
\usepackage{tabularx}
\usepackage{pslatex}

\newcommand{\vn}{\boldsymbol{\nabla}}
\newcommand{\vE}{\mathbf{E}}

\newcommand{\vecb}[1]{\mathbf{#1}}

\renewcommand{\bottomfraction}{0.7}

\begin{document}

\title{Mode coupling and cavity-quantum-dot interactions in a fiber-coupled microdisk cavity}

\author{Kartik Srinivasan$^{1}$ and Oskar Painter$^{2}$}
\affiliation{$^{1}$Center for the Physics of Information,
California Institute of Technology, Pasadena, California 91125}
\affiliation{$^{2}$Department of Applied Physics, California
Institute of Technology, Pasadena, California 91125}
\email{kartik@caltech.edu}
\date{\today}
\begin{abstract}

  A quantum master equation model for the interaction between a two-level system and whispering-gallery modes (WGMs) of
  a microdisk cavity is presented, with specific attention paid to current experiments involving a semiconductor quantum
  dot (QD) embedded in a fiber-coupled, AlGaAs microdisk cavity.  In standard single mode cavity QED, three important
  rates characterize the system: the QD-cavity coupling rate $g$, the cavity decay rate $\kappa$, and the QD dephasing
  rate $\gamma_{\perp}$.  A more accurate model of the microdisk cavity includes two additional features.  The first is
  a second cavity mode that can couple to the QD, which for an ideal microdisk corresponds to a traveling wave WGM
  propagating counter to the first WGM. The second feature is a coupling between these two traveling wave WGMs, at a
  rate $\beta$, due to backscattering caused by surface roughness that is present in fabricated devices.  We consider
  the transmitted and reflected signals from the cavity for different parameter regimes of
  $\{g,\beta,\kappa,\gamma_{\perp}\}$.  A result of this analysis is that even in the presence of negligible
  roughness-induced backscattering, a strongly coupled QD mediates coupling between the traveling wave WGMs, resulting
  in an enhanced effective coherent coupling rate $g=\sqrt{2}g_{0}$ corresponding to that of a standing wave WGM with
  an electric field maximum at the position of the QD. In addition, analysis of the second-order correlation function of the
  reflected signal from the cavity indicates that regions of strong photon antibunching or bunching may be present
  depending upon the strength of coupling of the QD to each of the cavity modes.  Such intensity correlation information will
  likely be valuable in interpreting experimental measurements of a strongly-coupled QD to a bi-modal WGM cavity.
\end{abstract}


\maketitle

\setcounter{page}{1}
\section{Introduction}
\label{sec:intro}

Recent demonstrations of vacuum Rabi splitting in systems consisting of a semiconductor microcavity and a single quantum
dot (QD) \cite{ref:Reithmaier,ref:Yoshie3,ref:Peter} represent an important milestone in investigations of cavity QED in
solid-state materials. In these experiments, the microcavity-QD system is incoherently pumped with an excitation beam at
an energy above the bandgap of both the QD and surrounding semiconductor material (usually GaAs or some form of its
alloy AlGaAs).  This pump light is absorbed and generates carriers in the GaAs system that can eventually (through
phonon and carrier scattering) fill the states of the QD; under weak enough pumping conditions, only the lowest energy
bound exciton state of the QD is appreciably populated on average.  Radiative recombination of the exciton state and the
resulting spontaneous emission is then modified by the presence of a resonant microcavity.  When the cavity is of small
enough volume, the coupling ($g$) between the QD exciton and the cavity can be large, and if the cavity decay rate
$\kappa$ and QD dephasing rate $\gamma_{\perp}$ are smaller than $g$, the system is said to be strongly coupled
\cite{ref:Kimble2}, in that the QD exciton and cavity mode are no longer truly separate entities but are instead bound
together.  In the experiments described in Refs.  \cite{ref:Reithmaier,ref:Yoshie3,ref:Peter}, the evidence of this
strong coupling has been presented in the form of spontaneous emission measurements from the QD-microcavity system,
which display a double-peaked structure, rather than the single peak associated with either the cavity mode or QD
exciton alone.  This vacuum Rabi splitting \cite{ref:Sanchez-Mondragon,ref:Agarwal} is one signature of the strong
coupling regime in cavity QED.

Applications of strongly coupled QD-microcavity systems to areas such as nonlinear optics and quantum information science
\cite{ref:Turchette,ref:Rice,ref:Savage,ref:Alsing,ref:Armen2} will also require an ability to effectively couple light
into and out of the microcavity-QD device.  That is, rather than measuring the spontaneous emission of the system alone,
it is also important to have access to the cavity's optical response (transmission or reflection).  This is true if, for
example, one wants to examine the effect of a coupled QD-cavity system on the propagation of a subsequent beam through
the cavity \cite{ref:Turchette,ref:Birnbaum}, or if one wants to use the phase of the emerging transmitted signal within
some type of logic gate \cite{ref:Duan2}.  Indeed, in most cavity QED experiments involving an atom coupled to a
Fabry-Perot cavity, it is the cavity's transmitted or reflected signal that is typically observed
\cite{ref:Thompson,ref:Hood,ref:Pinske,ref:Boca}.

Following demonstrations of coupling to $\emph{silica}$-based cavities such as microspheres\cite{ref:Knight,ref:Cai} and
microtoroids\cite{ref:Armani}, we have recently shown that optical fiber tapers\cite{ref:Birks,ref:Knight} are an
effective means to couple light into and out of wavelength-scale, semiconductor microcavities such as photonic
crystals\cite{ref:Srinivasan7,ref:Barclay7} and microdisks\cite{ref:Borselli,ref:Srinivasan9}.  In addition, we have
shown that microdisk cavities are extremely promising candidates for semiconductor cavity QED experiments, with recent
demonstrations of cavity quality factors ($Q$s) in excess of 10$^5$\cite{ref:Srinivasan9,ref:Srinivasan12} for devices
with a mode volume ($V_{\text{eff}}$) of $\sim2-6(\lambda/n)^3$.  These $Q$ values are significantly larger than those
utilized in Refs.  \cite{ref:Reithmaier,ref:Yoshie3,ref:Peter}, and as a result, the devices that we consider are poised
to operate well within the strong coupling regime, where multiple coherent interactions between the QD and photon can
occur.  It is envisioned that initial experiments in this fiber-coupled microcavity-QD system will examine vacuum-Rabi
splitting through measurements of the transmission spectrum past the cavity; such measurements will be directly
analogous to recent measurements of vacuum Rabi splitting from one-and-the-same atom in a Fabry-Perot cavity
\cite{ref:Boca}.

\begin{figure}[t]
\begin{center}
\epsfig{figure=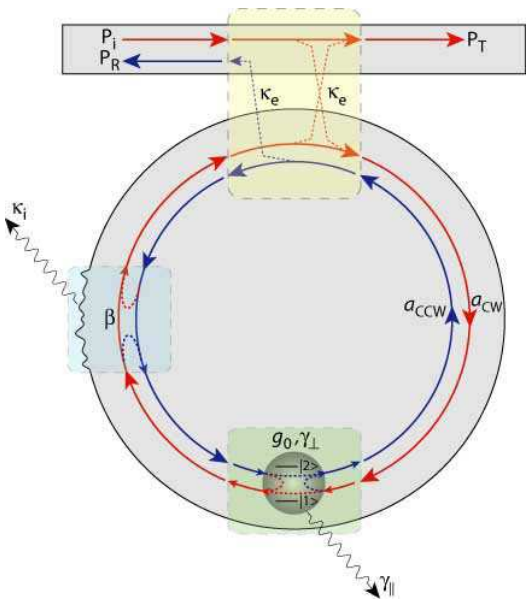, width=0.8\linewidth}
\caption{Illustration of the system under investigation. The
  microcavity-quantum-dot system is driven near resonance by coupling light into and out of it using an optical fiber
  taper waveguide, with a cavity-waveguide coupling rate $\kappa_{e}$ ($\kappa_{e}$ is a $\it{field}$ amplitude decay
  rate).  Imperfections in the microdisk cause a coupling of the clockwise and counterclockwise whispering-gallery
  modes, at a rate $\beta$.  These two whispering-gallery modes have a quantum-dot-cavity coupling rate $g_{0}$ and
  intrinsic cavity decay rate $\kappa_{i}$.  The quantum dot, approximated as a two-level system, has a radiative decay
  rate $\gamma_{\parallel}$ and a total transverse decay rate $\gamma_{\perp}$.} \label{fig:expt_configs}
\end{center}
\end{figure}

The goal of this paper is to provide a theoretical basis,
accompanied by numerical simulations, for the experiments to be
performed with single QDs in fiber-coupled microdisk cavities.  Of
particular concern is the proper treatment of the
whispering-gallery modes (WGMs) in the cavities.  More
specifically, the WGMs have a degeneracy of two as modes with
azimuthal number $\pm m$ have the same frequency, but circulate
around the disk in opposite directions.  The WGMs are typically
excited through an external waveguide, and for a nearly
phase-matched system the forward propagating mode through the
waveguide excites only the co-propagating mode in the resonator
(the clockwise (CW) traveling wave WGM from here on out).
Imperfections in the resonator will change this, as they cause
backscattering that can couple the CW and counterclockwise (CCW)
propagating modes (Fig. \ref{fig:expt_configs})
\cite{ref:Weiss,ref:Little3,ref:Kippenberg,ref:Borselli,ref:Borselli2}.
If the loss rates in the system (due to material absorption,
scattering loss, etc.) are low enough, the backscattering can lead
to coherent coupling of the CW and CCW modes, producing a pair of
standing wave modes. A similar theoretical model focused on cooled
alkali atoms coupled to dielectric whispering-gallery-mode
microcavities has been presented by Rosenblit, et al., in Ref.
\onlinecite{ref:Rosenblit}, and more recently by Aoki, et al., in
Ref. \onlinecite{ref:Aoki1}. In this work our interest is to study
this system in a parameter regime relevant to experiments
involving the interaction of a single self-assembled semiconductor
quantum dot with the microdisk WGMs in the presence of
roughness-induced backscattering \cite{ref:Srinivasan_thesis}, and
to determine the spectral response of the system for varying
degrees of quantum-dot-cavity coupling ($g_{0}$), backscattering
($\beta$), and modal loss ($\kappa_{T}$). We examine how the phase
and magnitude of the backscattering parameter affect the coupling
between one or both cavity modes and the QD, and how the QD itself
serves to couple the cavity modes together resulting in an
enhanced coherent coupling rate over that of traveling wave WGMs.

The organization of this paper is as follows: in section \ref{sec:modal_coupling}, we review the simple classical
coupled mode theory for modal coupling in microdisk cavities in absence of a QD. Section \ref{sec:QME_equations}
presents the quantum mechanical analysis of this system in the presence of a QD.  We review the quantum master equation
for this system and look at semiclassical approximations for specific choices of the backscattering parameter. In
section \ref{sec:QME_numerical_solns}, we present the results of numerical solutions of the quantum master equation for
parameters that are accessible in current experiments.  Finally, the intensity correlations in the reflected cavity
signal for various parameter regimes are studied in section \ref{sec:g2_calcs}.

\section{Modal coupling of two whispering-gallery modes due to surface scattering}
\label{sec:modal_coupling}

The modal coupling between CW and CCW traveling wave modes in a
whispering-gallery-mode microcavity has been observed
experimentally and explained by many other authors, including
those of Refs.
\onlinecite{ref:Weiss,ref:Little3,ref:Kippenberg,ref:Gorodetsky,ref:Borselli}.
Here, we present a simple analysis of this coupling.  This
analysis is essentially an abridged version of that which appears
in a recent paper by Borselli, et al., in Ref.
\onlinecite{ref:Borselli2}.

Maxwell's wave equation for the vector electric field in a
microdisk structure is

\begin{equation}
\label{eq:Maxwell_repeat}
\vn{^2}\vE-\mu_{0}\Bigl(\epsilon^{0}+\delta\epsilon\Bigr)\frac{\partial^{2}\vE}{\partial
t^{2}}=0,
\end{equation}

\noindent where $\mu_{0}$ is the permeability of free space,
$\epsilon^{0}$ is the dielectric function for the ideal (perfectly
cylindrical) microdisk and $\delta\epsilon$ is the dielectric
perturbation that is the source of mode coupling between the CW
and CCW modes. Assuming a harmonic time dependence, the
\emph{complex field} modes of the ideal ($\delta\epsilon=0$)
microdisk structure can be written as
$\vE^{0}_{j}(\vecb{r},t)=\vE^{0}_{j}(\vecb{r})\text{exp}(i\omega_{j}t)$,
where $j$ is an index label including the azimuthal number ($m$),
radial order ($p$), vertical order ($v$), and vertical parity (odd
or even for a cylinder with mirror symmetry).  In the microdisk
structures of interest the vertical height of the dielectric
cylinder is typically a half-wavelength in thickness, and only the
lowest order vertical mode is well localized to the microdisk. In
this case the vertical order and parity can be combined to define
the fundamental vertically guided whispering-gallery-modes of the
disk as transverse electric (TE), with antinode of the in-plane
($\hat{\rho},\hat{\phi}$) electric field components at the center
height of the disk, and transverse magnetic (TM), with antinode of
the vertical ($\hat{z}$) electric field component at the center
height of the disk.  In what follows we will continue to use the
TE and TM designation when discussing the WGMs, however, it should
be noted that due to the radial guiding of the modes in the small
microdisks of interest to this work the WGMs are far from actually
\emph{transverse} electric or magnetic, and contain significant
longitudinal field components in the azimuthal direction.

Solutions to eq. (\ref{eq:Maxwell_repeat}) with
$\delta\epsilon\neq0$ (i.e., modes of the perturbed structure) are written as a sum of the unperturbed mode basis

\begin{equation}
\label{eq:Efield_solns}
\vE(\vecb{r},t)=e^{-i\omega_{0}t}\sum_{j}a_{j}(t)\vE^{0}_{j}(\vecb{r}).
\end{equation}

\noindent Plugging into eq. (\ref{eq:Maxwell_repeat}),
keeping only terms up to first order, and utilizing mode
orthogonality, we arrive at a set of coupled mode equations

\begin{align}
\label{eq:coupled_mode_1}
&\frac{da_{k}}{dt}+i\Delta\omega_{k}a_{k}(t)=i\sum_{j}\beta_{jk}a_{j}(t) \\
&\beta_{jk}=\frac{\omega_{0}}{2}\frac{\int\delta\epsilon\bigl(\vE^{0}_{j}(\vecb{r})\bigr)^{\ast} \cdot \vE^{0}_{k}(\vecb{r})d\vecb{r}}{\int\epsilon^{0}|\vE^{0}_{k}(\vecb{r})|^{2}d\vecb{r}}.
\end{align}

Reference \cite{ref:Borselli2} presents a functional form for
$\beta$ in situations involving small surface roughness
perturbation.  Under weak scattering conditions an assumption is
made that only each pair (common radial order, etc.) of localized,
degenerate CW and CCW WGMs with azimuthal mode number $\pm m$ are
coupled by the disk perturbation $\delta\epsilon$.  The complex
electric fields of the CW and CCW WGMs are simply
related\cite{ref:Snyder_Love}, and can be written in a cylindrical
$(\rho,\phi,z)$ coordinate system as

\begin{equation}
\begin{split}
\label{eq:cw_ccw_complex_fields}
\vE^{0}_{CW}(\vecb{r})=(E^{0}_{\rho}(\rho,z),iE^{0}_{\phi}(\rho,z),E^{0}_{z}(\rho,z))e^{im\phi}, \\
\vE^{0}_{\text{CCW}}(\vecb{r})=(E^{0}_{\rho}(\rho,z),-iE^{0}_{\phi}(\rho,z),E^{0}_{z}(\rho,z))e^{-im\phi}.
\end{split}
\end{equation}

\noindent In the case of high-$Q$ resonant modes, with a small
degree of loss per round-trip, the CW and CCW WGMs are to a good
approximation complex conjugates of each other, which when
combined with eq. (\ref{eq:cw_ccw_complex_fields}) indicate that
the WGMs can be written with transverse ($\hat{\rho},\hat{z}$)
electric field components real and longitudinal ($\hat{\phi}$)
components imaginary\cite{ref:Snyder_Love} (i.e.,
$E^{0}_{\rho},E^{0}_{\phi},E^{0}_{z}$ all real functions).  The
coupled mode equations for these traveling wave modes then read as

\begin{equation}
\begin{split}
\label{eq:coupled_mode_2}
\frac{da_{\text{CW}}}{dt}&=-i\Delta{\omega}a_{\text{CW}}(t)+i|\beta|e^{i\xi}a_{\text{CCW}}(t), \\
\frac{da_{\text{CCW}}}{dt}&=-i\Delta{\omega}a_{\text{CCW}}(t)+i|\beta|e^{-i\xi}a_{\text{CW}}(t),
\end{split}
\end{equation}

\noindent with $\beta=|\beta|e^{i\xi}$ given by (in a basis with the transverse electric field components of the WGMs real),

\begin{equation}
\label{eq:backscatter_2}
\beta=\frac{\omega_{0}}{2}\frac{\int\bigl(\int \delta\epsilon\, e^{+i2m\phi} d\phi \bigr) \bigl((\vE^{0}_{\rho})^2-(\vE^{0}_{\phi})^2+(\vE^{0}_{z})^2\bigr) \rho d\rho dz}{2\pi\int\epsilon^{0}\bigl((\vE^{0}_{\rho})^2+(\vE^{0}_{\phi})^2+(\vE^{0}_{z})^2\bigr) \rho d\rho dz}.
\end{equation}

Equation (\ref{eq:coupled_mode_2}) represents the time evolution
of the two mode amplitudes ($a_{\text{CW}}$,$a_{\text{CCW}}$) of
an isolated system, without loss or coupling to an external
waveguide.  For the experiments considered in our work, the
waveguide coupler will be an optical fiber taper through which
light is traveling in the forward propagating mode. Light coupled
from the fiber taper will selectively excite the clockwise WGM of
the microdisk structure due to phase-matching.  Following the
formalism of Ref. \onlinecite{ref:Haus_Book} this
waveguide-resonator coupling can be included through the addition
of a term $k s_{+}$ to eq. (\ref{eq:coupled_mode_2}), where $k$ is
a waveguide coupling coefficient and $|s_{+}|^{2}$ is the input
{\emph{power}} in the external waveguide (the squared magnitude of
the mode amplitudes, $|a_{cw,ccw}|^2$, are normalized to stored
optical \emph{energy} in the cavity).  Loss is introduced to the
coupled mode equations by use of the phenomenological field
\emph{amplitude} decay rate $\kappa_{T}$, taken to be the same for
both the CW and CCW modes (though in general this does not have to
be the case).  This total field decay rate is broken into a
contribution from intrinsic microdisk absorption and scattering
loss ($\kappa_{i}$) and a contribution due to coupling to the
external waveguide ($\kappa_{e}$), so that
$\kappa_{T}=\kappa_{i}+\kappa_{e}$. Assuming lossless coupling and
time reversal symmetry it can be shown \cite{ref:Haus_Book} that
$|k|^{2}=2\kappa_{e}$. The coupled mode equations then read:

\begin{equation}
\begin{split}
\label{eq:coupled_mode_final}
\frac{da_{\text{CW}}}{dt}&=-\bigl(\kappa_{T}+i\Delta{\omega}\bigr)a_{\text{CW}}(t)+i|\beta|e^{i\xi}a_{\text{CCW}}(t)+\bigl(i\sqrt{2\kappa_{e}}\bigr)s_{+} \\
\frac{da_{\text{CCW}}}{dt}&=-\bigl(\kappa_{T}+i\Delta{\omega}\bigr)a_{\text{CCW}}(t)+i|\beta|e^{-i\xi}a_{\text{CW}}(t),
\end{split}
\end{equation}

\noindent Here, the phase of the coupling coefficient was
(arbitrarily) chosen to be purely imaginary, corresponding to a
single-pass, waveguide transmission coefficient of $+1$ in the
lossless coupler case\cite{ref:Haus_Book}.  These two coupled
equations can be rewritten as uncoupled equations in terms of the
variables $a_{\text{SW,1}}$ and $a_{\text{SW,2}}$, which represent
the standing wave mode amplitudes

\begin{equation}
\begin{split}
\label{eq:standing_wave_modes_repeat}
a_{\text{SW,1}}&=\frac{1}{\sqrt{2}}\bigl(a_{\text{CW}}+e^{i\xi}a_{\text{CCW}}\bigr) \\
a_{\text{SW,2}}&=\frac{1}{\sqrt{2}}\bigl(a_{\text{CW}}-e^{i\xi}a_{\text{CCW}}\bigr).
\end{split}
\end{equation}

\noindent As mentioned above, for an ideal microdisk the field
distributions of mode amplitudes $a_{\text{CW}}$ and
$a_{\text{CCW}}$ have an azimuthal spatial dependence of $e^{\pm
im\phi}$, so that the field distributions of $a_{\text{SW,1}}$ and
$a_{\text{SW,2}}$ correspond to (up to an overall phase factor)
standing waves $\sqrt{2}\cos(m\phi-\xi/2)$ and
$\sqrt{2}\sin(m\phi-\xi/2)$, respectively, with the azimuthal
orientation of the standing waves being determined by the phase
$\xi$ of the backscattering parameter.  Here, and in what follows,
we take the origin of the azimuthal axis ($\phi=0$) to be centered
at the QD.

\begin{figure*}[t]
\begin{center}
\epsfig{figure=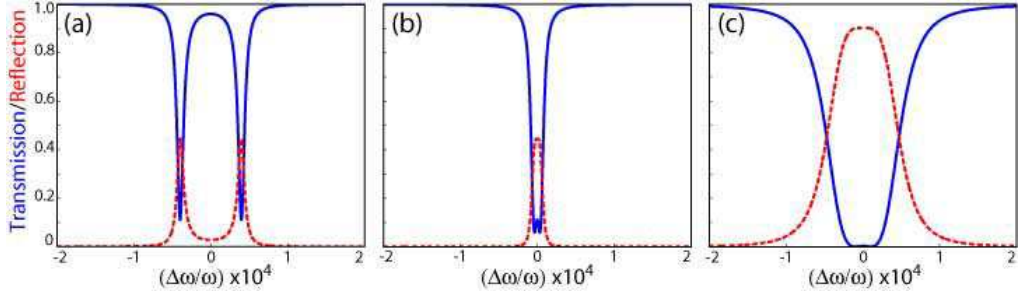, width=0.75\linewidth}
\caption{Normalized transmitted (solid line) and reflected (dashed
line) signal for standing wave whispering-gallery
  modes, determined through steady state solution of the coupled mode equations given in equation
  (\ref{eq:coupled_mode_final}). (a) $\beta/\kappa_{T}=8$, $\kappa_{T}/\kappa_{i}=3$ (b) $\beta/\kappa_{T}=1$,
  $\kappa_{T}/\kappa_{i}=3$, and (c) $\beta/\kappa_{T}=1$, $\kappa_{T}/\kappa_{i}=20$.  $Q_{i}=3{\times}10^5$ in all
  cases.}
\label{fig:doublet_model_classical_no_atom}
\end{center}
\end{figure*}

The transmitted ($P_{T}$) and reflected ($P_{R}$) optical power in
the external waveguide can be determined in either the basis of
$(a_{\text{CW}},a_{\text{CCW}})$ or
$(a_{\text{SW,1}},a_{\text{SW,2}})$; because of phase-matching,
coupling between the external waveguide and WGM resonator occur
directly through $(a_{\text{CW}},a_{\text{CCW}})$ and it is most
natural to solve for these quantities in the traveling wave mode
basis.  With the phase of the coupling coefficient chosen as
purely imaginary the transmitted and reflected powers are
$P_T=|s_{+} + \bigl(i\sqrt{2\kappa_{e}}\bigr)a_{\text{CW}}|^{2}$
and $P_R=|\bigl(i\sqrt{2\kappa_{e}}\bigr)a_{\text{CCW}}|^{2}$,
respectively. Steady state solutions for the normalized
transmitted and reflected signals from the cavity for a number of
different parameters are shown in Fig.
\ref{fig:doublet_model_classical_no_atom}.  For $\beta>\kappa_{T}$
(Fig. \ref{fig:doublet_model_classical_no_atom}(a)), we see the
formation of a distinct pair of resonances, located at
$\omega\approx\omega_{0}\pm\beta$.  These dips correspond to
standing wave resonances that result from a backscattering rate
($\beta$) that exceeds all other losses in the system
($\kappa_{T}$) so that coherent coupling between the CW and CCW
modes can take place. As we see in Fig.
\ref{fig:doublet_model_classical_no_atom}(b)-(c), for
$\beta\sim\kappa_{T}$, the resonances begin to overlap and are no
longer distinguishable.

For cavity QED applications, one very important consequence of the distinction between traveling wave and standing wave
modes is in the effective volume of the mode ($V_{\text{eff}}$), as the peak electric field strength per photon in the
cavity scales as $1/\sqrt{V_{\text{eff}}}$.  In particular, we recall the definition of $V_{\text{eff}}$ as:

\begin{equation}
\label{eq:V_eff}
V_{\text{eff}}=\frac{\int\epsilon|\vE(\vecb{r})|^{2}}{\max[\epsilon|\vE(\vecb{r})|^{2}]}.
\end{equation}

\noindent Standing wave WGMs have approximately half the volume of
the traveling wave WGMs, so that the coupling rate $g$ between a single quantum dot and a single photon in a standing
wave cavity mode is expected to be $\sqrt{2}$ times that when the quantum dot is coupled to a traveling wave cavity
mode.  This of course assumes the single QD is positioned at an antinode of the standing wave mode; alternately, if it
happens to be positioned at a node, the coupling rate $g$ will be zero.

These arguments again rely upon having a physical system in which
the backscattering coupling between CW and CCW modes is
sufficiently strong compared to all other loss rates to allow for
coherent modal coupling and formation of standing waves.  They
have also neglected the effects that an embedded QD may have, due
to both an introduction of additional loss and mode coupling into
the system.  In the case of a strongly coupled QD we might expect
that standing wave modes can be maintained provided that the modal
coupling rate $\beta$ exceeds not only $\kappa_{T}$ but also the
QD spontaneous emission rate $\gamma_{\parallel}$ and
non-radiative dephasing rate $\gamma_{p}$. To verify our physical
intuition and understand the system in better detail, we consider
a quantum master equation approach \cite{ref:Carmichael} to take
into account the cavity-QD interaction.

\section{Quantum master equation model}
\label{sec:QME_equations}

We begin by considering the Hamiltonian for an empty microdisk
cavity (traveling wave WGM resonance frequency $\omega_{c}$) with
field operators $\hat{a}_{\text{CW}}$ and $\hat{a}_{\text{CCW}}$
and mode coupling parameter $\beta$, written in a frame rotating
at the driving frequency $\omega_{l}$ (and for $\hbar=1$):

\begin{equation}
\begin{split}
\label{eq:empty_cavity_Hamiltonian}
H_{0}=&\Delta\omega_{cl}{\hat{a}_{\text{CW}}}^{\dagger}\hat{a}_{\text{CW}}+\Delta\omega_{cl}{\hat{a}_{\text{CCW}}}^{\dagger}\hat{a}_{\text{CCW}}-\beta{\hat{a}_{\text{CW}}}^{\dagger}\hat{a}_{\text{CCW}}\\
&-\beta^{\ast}{\hat{a}_{\text{CCW}}}^{\dagger}\hat{a}_{\text{CW}}+i(E{\hat{a}_{\text{CW}}}^{\dagger}-E^{\ast}\hat{a}_{\text{CW}}),
\end{split}
\end{equation}

\noindent  where $\Delta\omega_{cl}=\omega_{c}-\omega_{l}$.  As in
the coupled-mode equations of the previous section, the CW
propagating mode is driven by a classical intracavity field
$E=i\sqrt{2{\kappa_{e}}P_{in}}$, where $\kappa_{e}$ is the cavity
field decay rate into the waveguide and $P_{in}$ is the input
power in the external waveguide.  From this Hamiltonian, the
classical coupled-mode equations without dissipation can easily be
derived through an application of Ehrenfest's theorem.

Modeling the QD as a two-level system, we add the term $H_{1}$ to the Hamiltonian:

\begin{equation}
\begin{split}
\label{eq:atom_coupling_Hamiltonian}
H_{1}=&\Delta\omega_{al}\hat{\sigma}_{+}\hat{\sigma}_{-}+ig_{0}({\hat{a}_{\text{CW}}}^{\dagger}\hat{\sigma}_{-}-\hat{a}_{\text{CW}}\hat{\sigma}_{+})\\
&+ig_{0}({\hat{a}_{\text{CCW}}}^{\dagger}\hat{\sigma}_{-}-\hat{a}_{\text{CCW}}\hat{\sigma}_{+}),
\end{split}
\end{equation}

\noindent where $\Delta\omega_{al}=\omega_{a}-\omega_{l}$,
$\omega_{a}$ is the transition frequency of the exciton state of
the QD, and $g_{0}$ is the coherent coupling strength between the
QD exciton state and the \emph{traveling wave} WGMs.  Note that
$g_{0}$ has been assumed real, and to have the same phase for both
CW and CCW WGMs in eq. (\ref{eq:atom_coupling_Hamiltonian}). This
is consistent with a choice of the azimuthal origin lying at the
location of the QD and for a QD dipole polarization transverse to
the $\hat{\phi}$ direction, where the electric field components
for both WGMs are equal and real at the position of the QD (a WGM
basis can also be chosen in which this is true for dipole
polarization parallel to $\hat{\phi}$).  For a QD located away
from the azimuthal zero or with a mixed transverse and parallel
dipole orientation, $g_{0}$ will be complex, having a different
phase for the CW and CCW modes.  In general, care must be taken to
calculate $g_{0}$ and $\beta$ consistently when studying
interference effects between QD dipole scattering and
roughness-induced scattering.

The equation of motion
for the system's density matrix $\rho$ can be found from the equation:

\begin{equation}
\label{eq:density_matrix_no_loss}
\frac{d\rho}{dt}=\frac{1}{i}[H_{0}+H_{1},\rho]+L\rho
\end{equation}

\noindent where the term $L\rho=(L_{1}+L_{2}+L_{3})\rho$ allows
for the inclusion of decay through cavity loss (at a rate
$\kappa_{T}=\omega_{c}/2Q$), quantum dot spontaneous emission (at
a rate $\gamma_{\parallel}$), and phase-destroying collisional
processes (at a rate $\gamma_{p}$), which are of particular
importance for quantum dots, as unlike atoms, they are embedded in
a semiconductor matrix where electron-phonon scattering is
non-negligible.  In the zero-temperature limit (applicable to the
experiments under consideration as they will occur at cryogenic
temperatures), these loss terms are given by
\cite{ref:Carmichael,ref:Carmichael2}:

\begin{equation}
\label{eq:loss_terms_1}
\begin{split}
L_{1}\rho=&\kappa_{T}(2\hat{a}_{\text{CW}}\rho{\hat{a}_{\text{CW}}}^{\dagger}-{\hat{a}_{\text{CW}}}^{\dagger}\hat{a}_{\text{CW}}\rho-\rho{\hat{a}_{\text{CW}}}^{\dagger}\hat{a}_{\text{CW}}) \\
&+
\kappa_{T}(2\hat{a}_{\text{CCW}}\rho{\hat{a}_{\text{CCW}}}^{\dagger}-{\hat{a}_{\text{CCW}}}^{\dagger}\hat{a}_{\text{CCW}}\rho-\rho{\hat{a}_{\text{CCW}}}^{\dagger}\hat{a}_{\text{CCW}})
\end{split}
\end{equation}
\begin{equation}
\label{eq:loss_terms_2}
L_{2}\rho=\frac{\gamma_{\parallel}}{2}(2\hat{\sigma}_{-}\rho\hat{\sigma}_{+}-\hat{\sigma}_{+}\hat{\sigma}_{-}\rho-\rho\hat{\sigma}_{+}\hat{\sigma}_{-})
\end{equation}
\begin{equation}
\label{eq:loss_terms_3}
L_{3}\rho=\frac{\gamma_{p}}{2}(\hat{\sigma}_{z}\rho\hat{\sigma}_{z}-\rho)
\end{equation}

From the master equation, we can numerically calculate the steady
state density matrix $\rho_{ss}$ and relevant operator expectation
values such as
$\langle\hat{a}^{\dagger}_{\text{CW}}\hat{a}_{\text{CW}}\rangle_{ss}$,
which will then allow us to determine the transmission and
reflection spectrum of the coupled QD-cavity system using formulas
that are analogous to those used in the classical model of section
\ref{sec:modal_coupling}.  These calculations are the subject of
the following section.  For now, however, we consider what
intuition may be gained by further analytical study of the master
equation. Using the standing wave mode operators,

\begin{equation}
\begin{split}
\label{eq:standing_wave_mode_operators}
\hat{a}_{\text{SW,1}}&=\frac{1}{\sqrt{2}}\bigl(\hat{a}_{\text{CW}}+e^{i\xi}\hat{a}_{\text{CCW}}\bigr) \\
\hat{a}_{\text{SW,2}}&=\frac{1}{\sqrt{2}}\bigl(\hat{a}_{\text{CW}}-e^{i\xi}\hat{a}_{\text{CCW}}\bigr).
\end{split}
\end{equation}

\noindent and writing $\beta=|\beta|e^{i\xi}$, we take operator
expectation values to arrive at:

\begin{gather}
\label{eq:semiclassical_eqns_sw_basis}
\begin{split}
&\frac{d}{dt}\langle\hat{a}_{\text{SW,1}}\rangle=-i\bigl(\Delta\omega_{cl}-|\beta|\bigr)\langle\hat{a}_{\text{SW,1}}\rangle\\
&\qquad \qquad \quad +g_{0}\frac{1+e^{i\xi}}{\sqrt{2}}\langle\hat{\sigma}_{-}\rangle-\kappa_{T}\langle\hat{a}_{\text{SW,1}}\rangle+\frac{E}{\sqrt{2}}\\
&\frac{d}{dt}\langle\hat{a}_{\text{SW,2}}\rangle=-i\bigl(\Delta\omega_{cl}+|\beta|\bigr)\langle\hat{a}_{\text{SW,2}}\rangle\\
&\qquad \qquad \quad +g_{0}\frac{1-e^{i\xi}}{\sqrt{2}}\langle\hat{\sigma}_{-}\rangle-\kappa_{T}\langle\hat{a}_{\text{SW,2}}\rangle+\frac{E}{\sqrt{2}}\\
&\frac{d}{dt}\langle\hat{\sigma}_{-}\rangle=-\bigl(i\Delta\omega_{al}+\gamma_{\perp}\bigr)\langle\hat{\sigma}_{-}\rangle+\frac{g_{0}}{\sqrt{2}}\Bigl(\langle\hat{\sigma}_{z}\hat{a}_{\text{SW,1}}\rangle\bigl(1+e^{-i\xi}\bigr)\\
&\qquad \qquad +\langle\hat{\sigma}_{z}\hat{a}_{\text{SW,2}}\rangle\bigl(1-e^{-i\xi}\bigr)\Bigr)\\
&\frac{d}{dt}\langle\hat{\sigma}_{z}\rangle=-\sqrt{2}g_{0}\Bigl(\langle{\hat{\sigma}_{-}\hat{a}_{\text{SW,1}}}^{\dagger}\rangle\bigl(1+e^{i\xi}\bigr)+\langle{\hat{\sigma}_{-}\hat{a}_{\text{SW,2}}}^{\dagger}\rangle\bigl(1-e^{i\xi}\bigr)\Bigr)\\
&\qquad \qquad -\sqrt{2}g_{0}\Bigl(\langle\hat{\sigma}_{+}\hat{a}_{\text{SW,1}}\rangle\bigl(1+e^{-i\xi}\bigr)+\langle{\hat{\sigma}_{+}\hat{a}_{\text{SW,2}}}\rangle\bigl(1-e^{-i\xi}\bigr)\Bigr)\\
&\qquad\qquad-\gamma_{\parallel}\bigl(1+\langle\hat{\sigma}_{z}\rangle\bigr).
\end{split}
\end{gather}

\noindent where we have used
$[\hat{\sigma}_{+},\hat{\sigma}_{-}]=\hat{\sigma}_{z}$ and
$\gamma_{\perp}=\gamma_{\parallel}/2+\gamma_{p}$. In the new
standing wave mode basis both the empty-cavity frequencies and the
QD-cavity coupling strengths are seen to be modified by the
presence of strong backscattering.  For the low-frequency mode
($\omega_{c}-|\beta|$) corresponding to field operator
$\hat{a}_{\text{SW,1}}$, the effective coupling strength is
$g_{\text{SW,1}}=g_{0}(1+e^{i\xi})/\sqrt{2}$, while for the
high-frequency mode ($\omega_{c}+|\beta|$) corresponding to field
operator $\hat{a}_{\text{SW,2}}$, the effective coupling strength
is $g_{\text{SW,2}}=g_{0}(1-e^{i\xi})/\sqrt{2}$.  These coupling
strengths are thus dependent on the phase $\xi$ of the
backscattering parameter $\beta$ and can be as large as
$\sqrt{2}g_{0}$ or as small as zero.  This result is consistent
with what one would expect intuitively; the superposition of
traveling wave modes results in a pair of standing wave modes
whose peak field strength (per photon) is $\sqrt{2}$ times that of
a traveling wave mode. The two standing wave modes are phase
shifted from each other in the azimuthal direction by $\pi/2$, and
as a result, if the QD is positioned in the antinode of one mode
($\xi$=0, so that $g_{\text{SW,1}}=\sqrt{2}g_{0}$), it is within a
node of the other mode (so that $g_{\text{SW,2}}=0$), and vice
versa for the situation when $\xi$=$\pi$.  Of course, for large
cooperativity, $C$, between the traveling wave cavity modes and
the QD, the modes ($\hat{a}_{\text{SW,1}},\hat{a}_{\text{SW,2}}$)
may no longer be a good eigenbasis of the system.  In order to
gain some insight into such situations before moving on to
numerical quantum master equation simulations, we consider below
steady state solutions to the semiclassical equations.

The semiclassical equations of motion, derived from the above
equations by assuming that expectation values of products of
operators equal the product of the expectation values, can be
solved in steady state to yield information about the cavity
response as a function of drive strength and detunings, and are
useful for understanding the linear and nonlinear spectroscopy of
the
system\cite{ref:Kimble,ref:Lugiato,ref:Thompson2,ref:Savage,ref:Armen2}.
In the case of a single cavity mode coupled to a two-level system
this leads to the standard optical bistability state equation
(OBSE).  We consider two examples from the microdisk model
described above with two cavity modes, one in which the scattering
by the QD and the roughness-induced backscattering are in phase
and the coupling between the QD and cavity mode are described by
the simple intuitive picture above, the other in which the two
processes compete and the system response is more complicated.  We
begin with the simplest case in which $\xi$=0 (the $\xi=\pi$ case
is identical except the roles of $\hat{a}_{\text{SW,1}}$ and
$\hat{a}_{\text{SW,2}}$ are swapped). Defining the parameters,

\begin{equation}
\begin{split}
\label{eq:OBSE_params_ch8}
&n_{s}=\frac{\gamma_{\perp}\gamma_{\parallel}}{4g_{0}^2}, \,\,C=\frac{g_{0}^2}{2\kappa_{T}\gamma_{\perp}}, \\
&Y=\frac{E}{\sqrt{2n_{s}}\kappa_{T}}, \\
&X_{+}=\frac{\langle\hat{a}_{\text{SW,1}}\rangle|_{\xi=0}}{\sqrt{n_{s}}}, \,\,X_{-}=\frac{\langle\hat{a}_{\text{SW,2}}\rangle|_{\xi=0}}{\sqrt{n_{s}}}, \\ \\ \\ \\ \\
\end{split}
\end{equation}

\noindent and solving eq. (\ref{eq:semiclassical_eqns_sw_basis})
with $\xi=0$ in steady state we arrive at the following
expressions relating the external drive ($Y$) to the internal
state of the cavity ($X_{+},X_{-}$):

\begin{widetext}
\begin{center}
\begin{equation}
\begin{split}
\label{eq:OBSE_xi_0}
X_{+}&=\cfrac{Y}{1+\frac{4C}{2|X_{+}|^2+\bigl(\frac{\Delta\omega_{al}}{\gamma_{\perp}}\bigr)^2+1}+i\Biggl(\frac{\Delta\omega_{cl}-\beta}{\kappa_{T}}-\frac{4C\bigl(\frac{\Delta\omega_{al}}{\gamma_{\perp}}\bigr)}{2|X_{+}|^2+\bigl(\frac{\Delta\omega_{al}}{\gamma_{\perp}}\bigr)^2+1}\Biggr)}, \\
X_{-}&=\cfrac{Y}{1+i\Bigl(\frac{\Delta\omega_{cl}+\beta}{\kappa_{T}}\Bigr)}.
\end{split}
\end{equation}
\end{center}

\noindent Due to the common phase of the backscattering and the QD
mode coupling in this case, the net effect of the backscattering
on the system response is simply to shift the resonance
frequencies of the empty-cavity modes.  As expected the QD couples
to one standing wave mode with a cooperativity twice that of a
traveling wave mode, and is decoupled from the other.

In the general case both standing wave WGMs couple to the QD and
obtaining an equation analogous to the OBSE for an arbitrary $\xi$
is somewhat algebraically tedious.  As a simple example in which
both modes are coupled to the QD we consider $\xi=\pi/2$, which
yields in steady-state:

\begin{center}
\begin{equation}
\begin{split}
\label{eq:OBSE_xi_pi_2}
X_{+}&=\frac{Y}{\Bigl(\frac{1+i\Delta\omega_{cl}/\kappa_{T}}{1+i\Delta\omega_{cl}/\kappa_{T}+|\beta|/\kappa_{T}}\Bigr)\Biggl[
1+\frac{(|\beta|/\kappa_{T})^2}{1+\bigl(\frac{\Delta\omega_{cl}}{\kappa_{T}}\bigr)^2} + \frac{4C}{2X_{+}^2+\bigl(\frac{\Delta\omega_{al}}{\gamma_{\perp}}\bigr)^2+1} + i\Biggl(\frac{\Delta\omega_{cl}}{\kappa_{T}}\biggl(1-\frac{(|\beta|/\kappa_{T})^2}{1+\bigl(\frac{\Delta\omega_{cl}}{\kappa_{T}}\bigr)^2}\biggr) - \frac{4C\bigl(\frac{\Delta\omega_{al}}{\gamma_{\perp}}\bigr)}{2X_{+}^2+\bigl(\frac{\Delta\omega_{al}}{\gamma_{\perp}}\bigr)^2+1}  \Biggr)
\Biggr]}, \\
X_{-}&=\frac{Y -
\frac{|\beta|}{\kappa_{T}}X_{+}}{1+i\frac{\Delta\omega_{cl}}{\kappa_{T}}}.
\end{split}
\end{equation}
\end{center}
\end{widetext}

\noindent In this case the backscattering and the QD mode coupling are out
of phase and in competition.  The resulting system response is
governed by the detunings ($\Delta\omega_{al},\Delta\omega_{cl}$)
and the relative magnitude of the normalized roughness-induced
backscattering, $(|\beta|/\kappa_{T})^2$, and the cooperativity,
$C$.

Finally note that in order to connect to experiment an
input-output expression between the incoming optical signal in the
waveguide and the optical transmission (or reflection) past the
cavity into our collection fiber is desired.  In the formalism
presented in section \ref{sec:modal_coupling} the transmission and
reflection are given in terms of the traveling wave mode
amplitudes. These amplitudes can easily be recovered from even and
odd parity superpositions of $X_{+}$ and $X_{-}$ (cf. eq.
(\ref{eq:standing_wave_mode_operators})).

\begin{figure*}
\begin{center}
  \epsfig{figure=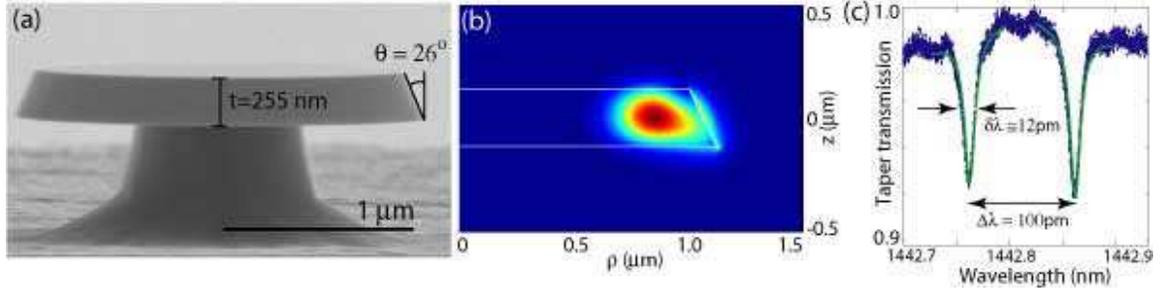, width=0.85\linewidth} \caption{(a) Scanning electron microscope (SEM) image of a
    fabricated microdisk device. The disk thickness is $t$=255 nm and the sidewall angle is $\theta=26^{\circ}$ from
    vertical.  The measured average diameter for this device (i.e., the diameter at the center of the slab) is $2.12$ $\mu$m. (b) Finite-element-calculated $|\vecb{E}|^2$ distribution for the TE$_{p=1,m=11}$ WGM of a microdisk
    with a diameter of $2.12$ $\mu$m at the center of the slab.  For this mode, $\lambda\sim1265.41$ nm,
    $Q_{\text{rad}}\sim10^7$, and for a traveling wave mode, $V_{\text{eff}}\sim5.6 (\lambda/n)^3$. (c) Typical measured normalized optical transmission spectrum of the TE$_{p=1,9}$ WGMs of a $2$ $\mu$m diameter microdisk similar to that in the SEM image of (a).} \label{fig:SEM_plus_FEMLAB}
\end{center}
\end{figure*}

\section{Solutions to the steady state quantum master equation in the weak driving regime}
\label{sec:QME_numerical_solns}

The quantum master equation (QME) presented in the previous section is solved numerically using the Quantum Optics
Toolbox \cite{ref:Tan1,ref:Tan2}.  We begin by considering steady state solutions, and calculate the
transmitted and reflected optical signals from the cavity.  As a starting point, we eliminate the
quantum dot from the problem by taking the coupling rate $g_{0}=0$. As expected, the resulting solutions (not displayed
here) are identical to those obtained using the classical coupled mode equations and presented in Fig.
\ref{fig:doublet_model_classical_no_atom}.  Having confirmed that the QME solution is consistent with the classical
solution in the empty cavity limit, we move on to study interactions with the quantum dot.  To connect these simulations
to ongoing experiments we choose physical parameters consistent with our fabricated devices
\cite{ref:Srinivasan9,ref:Srinivasan12}. In these experiments the microdisk cavity is 255 nm thick, and has a sidewall
angle of $26^{\circ}$ as shown in Figure \ref{fig:SEM_plus_FEMLAB}(a).  The modes of these structures (Fig.
\ref{fig:SEM_plus_FEMLAB}(b)) can be numerically investigated through finite-element eigenfrequency calculations using
the Comsol FEMLAB software \cite{ref:Spillane3,ref:Borselli3,ref:Srinivasan12}, and information about the effective
modal volume $V_{\text{eff}}$ (as defined in eq. (\ref{eq:V_eff})) and radiation-limited quality factor $Q_{\text{rad}}$ can be
obtained.  For the purposes of this work we focus on modes of transverse electric (TE) polarization, where the electric
field lies predominantly within the plane of the disk, and we consider first order radial modes ($p=1$) in the $1200$ nm
wavelength band, the wavelength region of the ground state exciton transition in our QDs.

\begin{figure}[b]
\begin{center}
  \epsfig{figure=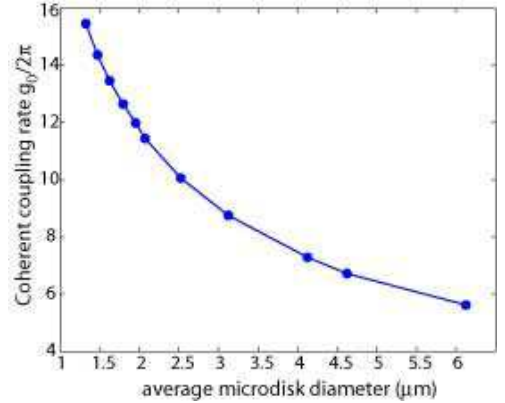, width=0.75\linewidth} \caption{Coherent coupling rate
  $g_{0}/2\pi$ (with $\tau_{\text{sp}}=1$ ns) for traveling wave TE$_{p=1,m}$
  whispering-gallery modes of the microdisk structure described in Fig. \ref{fig:SEM_plus_FEMLAB}
  with varying disk diameter.  Calculations were performed using a fully vectorial finite-element method,
  where for each microdisk diameter the azimuthal number of the TE$_{p=1,m}$ WGM resonance was adjusted to
  place the resonance frequency nearest $\lambda=1250$ nm.} \label{fig:sim_results_2_ch8}
\end{center}
\end{figure}

As discussed in ref. \cite{ref:Srinivasan12}, finite-element
method simulations can be used to calculate $V_{\text{eff}}$ as a
function of the average microdisk diameter $D_{\text{avg}}$. From
$V_{\text{eff}}$, we can estimate the QD-photon coupling strength.
For a QD located at a position of maximum electric field energy
density and with exciton dipole parallel to the local electric
field of the cavity mode,
$g_{0}=\vecb{d}\cdot\vecb{E_{\text{ph}}}/\hbar$ is given
by\cite{ref:Kimble2,ref:Andreani}

\begin{equation}
\label{eq:coupling_rate_repeat}
\begin{split}
g_{0}=\frac{1}{2\tau_{\text{sp}}}\sqrt{\frac{3c\lambda_{0}^2\tau_{\text{sp}}}{2\pi{n^3}V_{\text{eff}}}}, \\
\end{split}
\end{equation}

\begin{figure*}
\begin{center}
\epsfig{figure=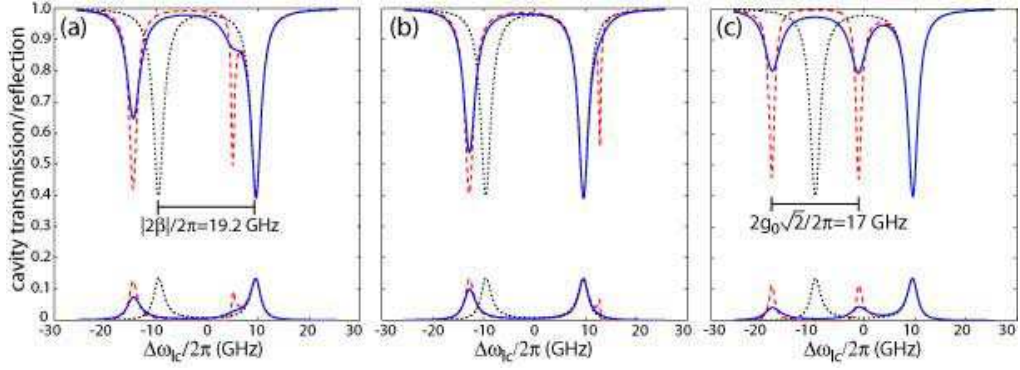, width=0.75\linewidth} \caption{Steady
state QME solution for the normalized optical transmission (top
curves) and reflection (bottom curves) spectra for a QD coupled to
a microdisk cavity under weak driving and for three different QD
detunings: (a)
  $\Delta\omega_{ac}=0$, (b) $\Delta\omega_{ac}=\beta$, and (c) $\Delta\omega_{ac}=-\beta$. Cavity and QD parameters for these simulations are
  $\{g_{0},\beta,\kappa_{T},\kappa_{e},\gamma_{\parallel},\gamma_p\}/2\pi=\{6,9.6,1.2,0.44,0.16,2.4\}$ GHz, with the phase of the backscattering parameter set to $\xi=0$.  In these plots the additional black dotted line plots correspond to an empty cavity ($g_{0}=0$) and the red dashed line plots
  correspond to a QD with no non-radiative dephasing ($\gamma_{p}/2\pi=0$ GHz).}
\label{fig:QME_sim_results_1}
\end{center}
\end{figure*}

\begin{figure*}
\begin{center}
\begin{minipage}[c]{0.55\linewidth}
\centering \epsfig{figure=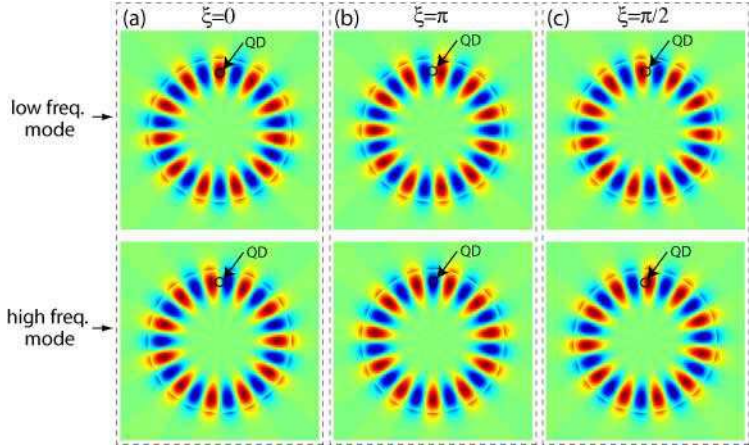, width=\linewidth}
\end{minipage}\hfill
\begin{minipage}[c]{0.36\linewidth}
\centering \caption{Standing wave modes in a microdisk for
different phases of $\beta$ showing how the low and high frequency
modes are positioned with respect to a fixed QD. (a) $\beta>0$
($\xi=0$), (b) $\beta<0$ ($\xi=\pi$), and (c) $\beta=i|\beta|$
($\xi=\pi/2$).\label{fig:udisk_modes_QD_diff_betas}}
\end{minipage}
\end{center}
\end{figure*}

\noindent where $\tau_{\text{sp}}$ is the spontaneous emission
lifetime of the QD exciton.  Consistent with what has been
measured experimentally for self-assembled InAs quantum
dots\cite{ref:Bayer}, we take $\tau_{\text{sp}} = 1$ ns.  Figure
\ref{fig:sim_results_2_ch8} shows a plot of $g_{0}$ versus disk
size for traveling wave WGMs, and we see that $g_{0}/2\pi$ can be
as high as 16 GHz for the range of diameters we consider.  As
discussed in ref. \cite{ref:Srinivasan12}, the WGMs are well
confined ($Q_{\text{rad}}>10^5$) for all but the smallest diameter
disks ($D_{\text{avg}}<1.5$ $\mu$m). We have confirmed this in
experiments\cite{ref:Srinivasan11,ref:Srinivasan12}, with $Q$ as
high as $3.6{\times}10^5$ measured, so that cavity decay rates
$\kappa_{T}/2\pi$ of approximately $1$ GHz can reasonably be
expected. Such devices exhibited doublet splittings that are on
the order of $\Delta\lambda=10$-$100$ pm (see Fig.
\ref{fig:SEM_plus_FEMLAB}(c)), corresponding to a backscattering
rate $|\beta|/2\pi=1$-$10$ GHz. In practical devices then, the
roughness-induced backscattering and the coherent QD-cavity mode
coupling rates can be of similar magnitude, and we thus expect the
QME simulation results to be particularly helpful in interpreting
future experimental data.

Unless otherwise specified, in all of the QME simulations to
follow we consider the weak driving limit.  In this limit the
steady state response of the system behaves linearly, with the
internal cavity photon number $\ll 1$ and QD saturation effects
negligible. For the QD and cavity parameters of the microdisk
structures described below this corresponds to input powers of
about $10$ pW.

\subsection{$\beta>g_{0}>(\kappa_{T},\gamma_{\perp})$}
\label{sec:QME_sims_1}

The first situation we study is one in which the backscattering rate $\beta$ exceeds the coupling rate $g_{0}$, which in
turn exceeds the cavity and QD decay rates $\kappa_{T}$ and $\gamma_{\perp}$.  We choose $\beta/2\pi=9.6$ GHz ($\xi=0$),
with $g_{0}/2\pi=6$ GHz, $\kappa_{T}/2\pi=1.2$ GHz (corresponding to $Q=10^5$), $\kappa_{e}/2\pi=0.44$ GHz
(corresponding to a transmission depth of $60\%$ for the empty-cavity standing wave modes), and $\tau_{\text{sp}}=1$ ns
($\gamma_{\parallel}/2\pi\sim0.16$ GHz).  The unperturbed cavity frequency (i.e., the resonance frequency of the
\emph{traveling wave} modes) is fixed at $\omega_{c}=0$, and three different QD-cavity detunings,
$\Delta\omega_{ac}=\omega_{a}-\omega_{c}$=$\{0,\beta,-\beta\}$ are considered.  For each value of $\Delta\omega_{ac}$,
we calculate the steady state transmission and reflection spectra (as a function of probe laser frequency to cavity
detuning, $\Delta\omega_{lc}=\omega_{l}-\omega_{c}$) from the cavity in three different limits: (i) $g_{0}$=0; here, there is no QD-cavity
coupling, and the response should be that of an empty cavity, (ii) $g_{0}/2\pi$=6 GHz, $\gamma_{p}/2\pi$=0 GHz; here, we
neglect all non-radiative dephasing, which becomes a good approximation as the temperature of the QD is cooled below
$10$ K, and (iii) $g_{0}/2\pi$=6 GHz, $\gamma_{p}/2\pi$=2.4 GHz; here, we allow for a significant amount of
non-radiative dephasing, corresponding to a QD exciton linewidth of $10$ $\mu$eV, which is consistent with what has
been observed experimentally at temperatures of around $10$-$20$ K \cite{ref:Bayer}.

\begin{figure*}[t]
\begin{center}
  \epsfig{figure=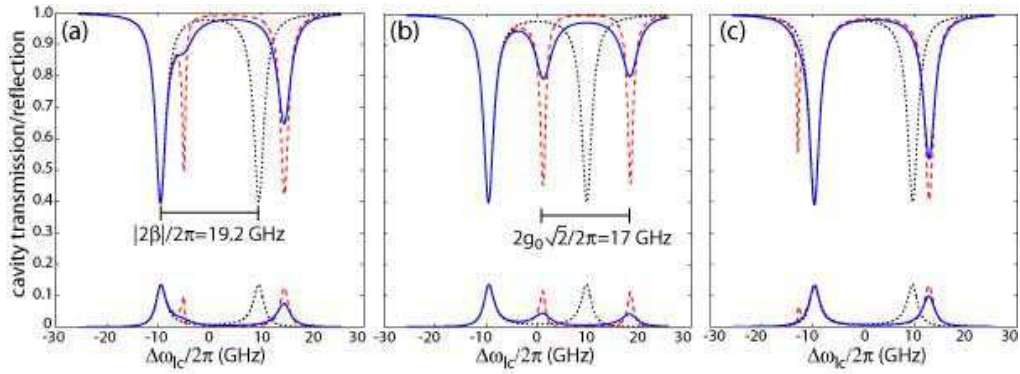, width=0.75\linewidth}
\caption{Steady state QME solution for the normalized optical
transmission (top curves) and reflection (bottom curves) spectra
of a QD coupled to a microdisk cavity under weak driving and for
three different QD detunings: (a)
  $\Delta\omega_{ac}=0$, (b) $\Delta\omega_{ac}=\beta$, and (c) $\Delta\omega_{ac}=-\beta$.  These plots are calculated for identical parameters as in Fig. \ref{fig:QME_sim_results_1} with the exception that the phase of the
  backscattering parameter $\beta$ has been changed from $\xi$=0 to $\xi$=$\pi$ ($\beta/2\pi$=-9.6 GHz).}
\label{fig:QME_sim_results_2}
\end{center}
\end{figure*}

\begin{figure*}
\begin{center}
\epsfig{figure=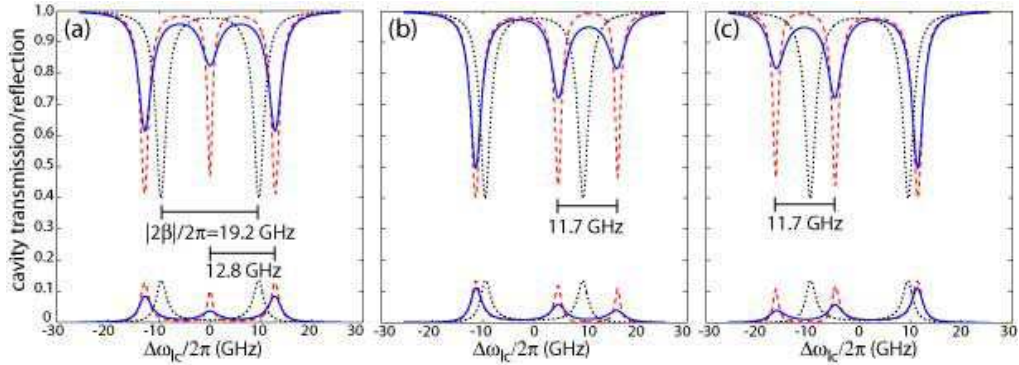, width=0.75\linewidth} \caption{Steady
state QME solution for the normalized optical transmission (top
curves) and reflection (bottom curves) spectra of a QD coupled to
a microdisk cavity under weak driving and for three different QD
detunings: (a)$\Delta\omega_{ac}=0$, (b)
$\Delta\omega_{ac}=\beta$, and (c) $\Delta\omega_{ac}=-\beta$.
These plots are calculated for
  identical parameters as those in Fig. \ref{fig:QME_sim_results_1} with the exception that the phase of the
  backscattering parameter $\beta$ has been changed from $\xi=0$ to $\xi=\pi/2$ ($\beta/2\pi=i9.6$ GHz).}
\label{fig:QME_sim_results_3}
\end{center}
\end{figure*}

\renewcommand{\bottomfraction}{0.3}

The results of the steady-state quantum master equation simulations are plotted in Fig. \ref{fig:QME_sim_results_1}. The
interpretation of these results is as follows: as a result of the modal coupling due to backscattering, which has formed
standing wave modes through a superposition of the initial traveling wave modes, only the lower frequency mode of the
doublet has any spatial overlap with the QD (see Fig.  \ref{fig:udisk_modes_QD_diff_betas} for location of the QD
relative to the two standing wave modes as a function of $\xi$), and thus, we should only expect the low frequency mode
to exhibit any frequency shifts or splittings.  In Fig. \ref{fig:QME_sim_results_1}(a), with the QD spectrally detuned
equally from both empty-cavity standing wave modes, we see \emph{asymmetric} vacuum Rabi splitting due to coupling of
the QD to the low frequency mode at $\omega_c-\beta$. In Fig.  \ref{fig:QME_sim_results_1}(b), with the QD now on
resonance with the higher frequency mode, coupling still only occurs to the low frequency mode detuned in this case by
$2\beta$.  Finally in Fig.  \ref{fig:QME_sim_results_1}(c), the QD is on resonance with the low frequency mode, and is
also spatially aligned with it, so that we see the familiar \emph{symmetric} vacuum Rabi splitting of this resonance.
We note that the frequency splitting, $\Omega_{R}$, is in this case $2\sqrt{2}g_{0}$ rather than $2g_{0}$; this is
consistent with the mode volume of the standing wave modes being one half that of the traveling wave modes.  For
$\xi=\pi$ (Fig. \ref{fig:QME_sim_results_2}) the results are the mirror image of those in Fig.
\ref{fig:QME_sim_results_1}, where now the high frequency mode is spatially aligned with the QD and exhibits frequency
shifts and vacuum Rabi splitting.

Finally, we consider an intermediate backscattering phase $\xi=\pi/2$.  Here, we expect both modes to have an equal (but
non-optimal) spatial alignment with the QD (Fig. \ref{fig:udisk_modes_QD_diff_betas}(c)).  The results, displayed in Fig.
\ref{fig:QME_sim_results_3}, show that this is indeed the case. In Fig. \ref{fig:QME_sim_results_3}(a), for example, we
see a symmetric spectrum, consistent with both modes being equally spatially coupled to the QD and equally (and
oppositely) spectrally detuned from it.  In Fig.  \ref{fig:QME_sim_results_3}(b)-(c), we see that the spectra are no
longer symmetric, as the QD is on resonance with the high frequency mode in Fig. \ref{fig:QME_sim_results_3}(b), and
with the low frequency mode in Fig. \ref{fig:QME_sim_results_3}(c).  In each case we see Rabi splitting about the mode
on resonance with the QD and only a small shift for the detuned mode.  The Rabi splitting between the peaks is no
longer at the maximum value of $2\sqrt{2}g_{0}$, but at a value closer to $2g_{0}$ due to the spatial misalignment of the QD with the empty-cavity standing wave modes.

\begin{figure}[t]
\begin{center}
\epsfig{figure=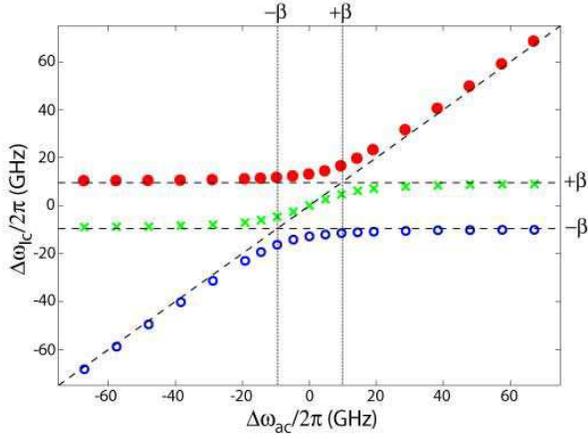,width=0.9\linewidth} \caption{Position
of the resonance dips within the transmission spectrum of a QD
coupled to a microdisk cavity under weak driving, as a function of
QD-cavity detuning $\Delta\omega_{ac}$.
$\{g_{0},\beta,\kappa_{T},\kappa_{e},\gamma_{\parallel},\gamma_p\}/2\pi=\{6,i9.6,1.2,0.44,0.16,2.4\}$
GHz, so that the QD is spatially coupled to both standing wave
cavity modes.} \label{fig:anti_crossing_diagram}
\end{center}
\end{figure}

Before moving on to study different parameter regimes for
$\{g_{0},\beta,\kappa,\gamma_{\perp}\}$, we examine the cavity's
transmission spectrum as a function of the spectral detuning of
the QD ($\Delta\omega_{ac}$). In practice
\cite{ref:Reithmaier,ref:Yoshie3,ref:Peter}, QD-cavity detuning is
often achieved by varying the sample temperature, which tunes at
different rates the transition frequency of the QD (due to its
temperature-dependent energy bandgap) and the cavity mode (due to
its temperature-dependent refractive index).  More recently, gas
condensation on the sample surface \cite{ref:Mosor} has been
successfully used to tune the cavity mode frequency of a
surface-sensitive photonic crystal microcavity.  Such a method hsa
recently been shown to be effective for the microdisks studied
here owing to the field localization at the top and bottom disk
surfaces and at the disk periphery\cite{ref:Srinivasan14}.  In
Fig. \ref{fig:anti_crossing_diagram} we plot the cavity
transmission minima as a function of $\Delta\omega_{ac}$ for the
parameter set studied above in Fig.  \ref{fig:QME_sim_results_3},
where the QD is spatially coupled to both standing wave modes of
the microdisk. When the QD is far detuned from the standing wave
cavity modes, we see the response of an essentially uncoupled
system, with transmission dips at the bare QD and cavity mode
frequencies ($\pm\beta$). In the center of the plot, as the QD is
tuned through the bare-cavity resonances, a pair of anti-crossings
are evident as the QD couples to each of the standing wave modes
of the microdisk.

\begin{figure*}
\begin{center}
\epsfig{figure=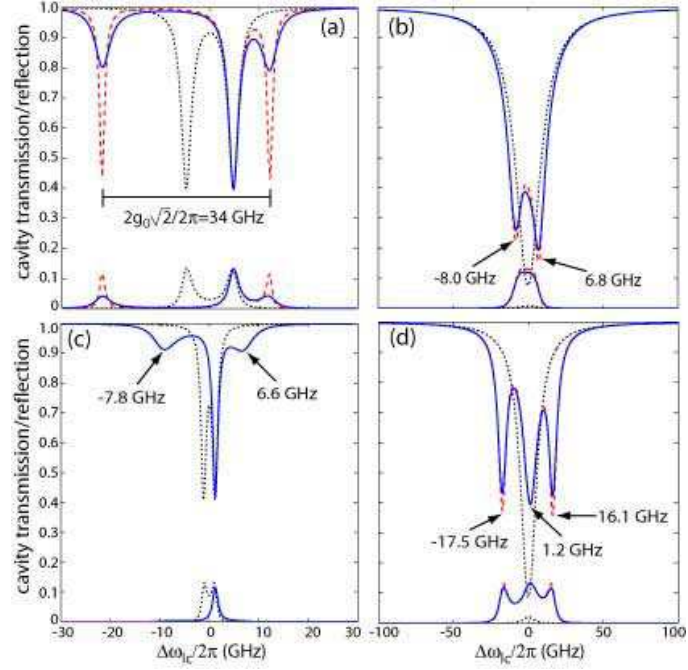,width=0.5\linewidth} \caption{Steady
state QME solution to the normalized optical transmission (top
curves) and reflection (bottom curves) spectra of a QD coupled
    to a microdisk cavity with $\xi=0$, and for: (a) $g_{0}>\beta>\kappa_{T}>\gamma_{\perp}$ ($\Delta\omega_{ac}=-\beta/2\pi$, $\{g_{0},\beta,\kappa_{T},\kappa_{e},\gamma_{\parallel},\gamma_{p}\}/2\pi=\{12,4.8,1.2,0.44,0.16,2.4\}$ GHz),
    (b) $\kappa_{T}>g_{0}>\beta>\gamma_{\perp}$ ($\Delta\omega_{ac}=0$, $\{g_{0},\beta,\kappa_{T},\kappa_{e},\gamma_{\parallel},\gamma_{p}\}/2\pi=\{6,1.2,9.6,3.5,0.16,0.7\}$ GHz), (c) $\gamma_{\parallel}>g_{0}>\beta>\kappa_{T}$ ($\Delta\omega_{ac}=-\beta/2\pi$, $\{g_{0},\beta,\kappa_{T},\kappa_{e},\gamma_{\parallel},\gamma_{p}\}/2\pi=\{6,1.2,0.6,0.22,9.4,0\}$ GHz), and (d)
    $g_{0}>\kappa_{T}>\beta>\gamma_{\perp}$ ($\Delta\omega_{ac}=0$, $\{g_{0},\beta,\kappa_{T},\kappa_{e},\gamma_{\parallel},\gamma_{p}\}/2\pi=\{12,1.2,6,2.2,0.16,0.7\}$ GHz).  In these plots the additional black dotted line plots correspond to an empty cavity ($g_{0}=0$) and the red dashed line plots
  correspond to a QD with no non-radiative dephasing ($\gamma_{p}/2\pi=0$ GHz).}
\label{fig:QME_sim_results_4}
\end{center}
\end{figure*}

\begin{figure}
\begin{center}
\epsfig{figure=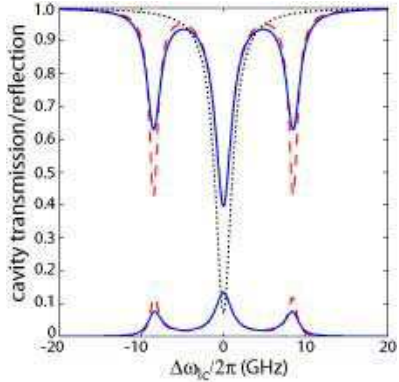,width=0.6\linewidth} \caption{Steady
state QME solution for the normalized optical transmission (top
curves) and reflection (bottom curves) spectra for a QD coupled to
a microdisk cavity under weak driving and with the
roughness-induced backscattering rate ($\beta$) zero.
$g_{0}>\kappa_{T}>\gamma_{\perp}>\beta$ ($\Delta\omega_{ac}=0$,
$\{g_{0},\beta,\kappa_{T},\kappa_{e},\gamma_{\parallel},\gamma_{p}\}/2\pi=\{6,0,1.2,0.44,0.16,0.7\}$
GHz).  The additional black dotted line plots correspond to an
empty cavity ($g_{0}=0$) and the red dashed lines plots
  correspond to a QD with no non-radiative dephasing ($\gamma_{p}/2\pi=0$ GHz).}
\label{fig:QME_solns_no_splitting}
\end{center}
\end{figure}

\subsection{$g_{0}>\beta>(\kappa_{T},\gamma_{\perp})$}
\label{sec:QME_sims_2}

Here we switch regimes to one in which the QD-cavity coupling rate dominates all other rates in the system,
including the backscattering rate $\beta$.  In particular, we choose $g_{0}/2\pi$=12 GHz, with $\beta/2\pi$=4.8 GHz,
$\kappa_{T}/2\pi$=1.2 GHz ($\kappa_{e}/2\pi$=0.44 GHz), and $\tau_{\text{sp}}$=1 ns ($\gamma_{\parallel}/2\pi\sim0.16$ GHz). The qualitative behavior
that we expect to see is similar to that of the previous section as both $g_{0}$ and $\beta$ represent coherent
processes, so that their relative values are not as important as their values in comparison to
the dissipative rates in the system. This is seen in Fig.  \ref{fig:QME_sim_results_4}(a), where the QD is spectrally
located at $-\beta$, so that it is resonant with the low frequency mode of the standing wave doublet.  Predictably, the
interaction with the QD causes this resonance to split, with a splitting $\Omega_{R}$=2$\sqrt{2}g_{0}$.  The higher
frequency mode remains unaffected, as the choice of $\xi$=0 causes it to be spatially misaligned with the QD.

\subsection{$\kappa_{T}>g_{0}>\beta>\gamma_{\perp}$}
\label{sec:QME_sims_3}

Now, we take the cavity loss rate $\kappa_{T}/2\pi$=9.6 GHz to exceed both $g_{0}/2\pi$=6 GHz and $\beta/2\pi$=1.2 GHz.
In addition, $\kappa_e/2\pi=3.5$ GHz, $\gamma_{\parallel}/2\pi$=0.16 GHz, and $\gamma_{p}/2\pi$=0 or 0.7 GHz, so that
$\kappa_{T}>\kappa_{e}>\gamma_\perp$ (good cavity limit). In the absence of a QD we expect to see a single transmission
dip rather than a doublet for $\kappa_{T}\gg\beta$.  This is confirmed in simulation by the black dotted line in Fig.
\ref{fig:QME_sim_results_4}(b).  With the addition of a QD, taken to be resonant with the center frequency of the single
cavity transmission dip, we expect to see this single dip split into two, with the dips not being completely resolved
due to decay of the cavity mode ($\kappa_{T}>g$).  This is confirmed in Fig.  \ref{fig:QME_sim_results_4}(b), where the
splitting $\Omega_{R}/2\pi=14.8$ GHz lies between the expected splitting for a purely traveling wave cavity mode
($\Omega_{R}$=2$g_{0}$) and the expected splitting for a purely standing wave cavity mode
($\Omega_{R}$=2$\sqrt{2}g_{0}$), and lies closer to the former due to the large degree to which $\kappa_{T}$ exceeds
$\beta$.

\subsection{$\gamma_{\parallel}>g_{0}>\beta>\kappa_{T}$}
\label{sec:QME_sims_4}

Here, the roles of $\kappa_{T}$ and $\gamma_{\parallel}$ are swapped in comparison to the previous subsection, so that
$\gamma_{\parallel}/2\pi$=9.6 GHz is the dominant dissipative rate, exceeding each of
$\{g_{0},\beta,\kappa_{T},\kappa_{e}\}/2\pi = \{6,1.2,0.6,0.22\}$ GHz (bad cavity limit).  Unlike our previous example,
in absence of a QD we do expect to see a pair of standing wave modes form, as $\beta>\kappa_{T}$. This is confirmed in
Fig.  \ref{fig:QME_sim_results_4}(c) (black dashed line). Now, we introduce a QD that is spectrally aligned with the low
frequency mode at $-\beta$. Because QD decay is so large in this case we expect that the standing wave character of the
modes is going to largely be erased when coupled to the QD. To confirm this intuition, we examine the calculated
transmission spectrum in Fig. \ref{fig:QME_sim_results_4}(c).  The low frequency mode does indeed split, but the
splitting $\Omega_{R}/2\pi=14.4$ GHz is less than the expected splitting of 2$\sqrt{2}g_{0}$ for standing wave
modes, and lies much closer to the 2$g_{0}$ splitting for traveling wave modes.  The situation thus mimics that of the
previous example, although in this case the relatively weak transmission contrast of the QD-coupled resonances is a result of operation in the bad cavity limit.

\subsection{$g_{0}>\kappa_{T}>\beta>\gamma_{\perp}$}
\label{sec:QME_sims_5}

Finally, we consider a scenario in which the QD-cavity coupling $g_{0}/2\pi=12$ GHz is the dominant rate in the
system, but where cavity decay $\kappa_{T}/2\pi$=6 GHz exceeds the backscattering rate $\beta/2\pi$=1.2 GHz.  In
absence of a QD we see a single transmission resonance dip (Fig. \ref{fig:QME_sim_results_4}(d)) as $\kappa_{T}>\beta$.
If a QD is now spectrally aligned to the center of this transmission dip ($\Delta\omega_{ac}$=0) three resonances
appear within the transmission spectrum of Fig. \ref{fig:QME_sim_results_4}(d).  This should be contrasted with the
transmission spectrum of Fig.  \ref{fig:QME_sim_results_4}(b) in which only two resonant transmission dips were present.
The central resonance dip of Fig. \ref{fig:QME_sim_results_4}(d) is at a detuned frequency of 1.2 GHz ($=\beta/2\pi$),
and corresponds to the frequency of one of the two standing wave modes that can form through an appropriate combination
of the traveling wave modes. As this mode is spatially misaligned from the QD for $\xi=0$, we do not expect its
frequency to have shifted due to interaction with the QD.  The other two transmission resonances correspond to the splitting of the
low frequency mode from its empty-cavity position at $-\beta/2\pi=-1.2$ GHz.  The splitting of $\Omega_{R}/2\pi$=33.6
GHz is very close to the value of 2$\sqrt{2}g_{0}$ expected for interaction with a standing wave mode.

\begin{figure*}[t]
\begin{center}
\epsfig{figure=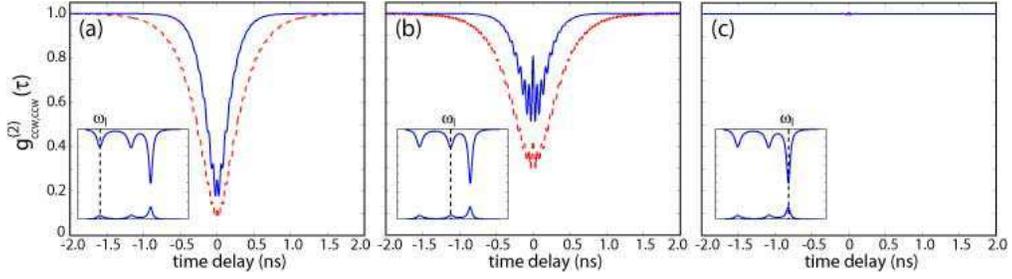, width=0.75\linewidth}
\caption{Normalized second order auto-correlation function (solid
blue line) $g^{(2)}_{ccw,ccw}(\tau)$ for the counterclockwise
propagating cavity mode for the parameters
$\{g_{0},\beta,\kappa_{T},\kappa_{e},\gamma_{\parallel},\gamma_p\}/2\pi=\{6,9.6,1.2,0.44,0.16,2.4\}$
GHz, $\Delta\omega_{ac}=-\beta$, and whose transmission/reflection
spectrum, originally shown in Fig. \ref{fig:QME_sim_results_1}(c),
is re-displayed here in the plot insets.
 (a) $\omega_{l}=-\beta-g_{0}\sqrt{2}$, (b) $\omega_{l}=-\beta+g_{0}\sqrt{2}$, and
 (c) $\omega_{l}=\beta$.  The additional red dashed line plots
correspond to a QD with no non-radiative dephasing
($\gamma_{p}/2\pi=0$ GHz).} \label{fig:QME_g2_tau_1}
\end{center}
\end{figure*}

The basic result that the above example demonstrates is that the
QD can effectively serve as a means to couple the traveling wave
microdisk modes, even in instances where the backscatter parameter
is small relative to other rates in the system.  As a final
illustration of this, we consider the situation where the
backscatter parameter is zero.  In Fig.
\ref{fig:QME_solns_no_splitting}, the empty-cavity single
transmission resonance separates into three resonance dips, one at
the original zero detuning and the other two split by
2$\sqrt{2}g_{0}$.  The interpretation of this result is that the
QD has effectively served to couple the two counter-propagating
traveling wave modes, creating a pair of standing wave resonant
modes, one which is decoupled and has an electric field node at
the position of the QD, and the other which is strongly coupled to
the QD at a field antinode.  In this case, and in other strong
coupling cases where $g_{0}$ is the dominant system rate, the QD
serves to set the position of the effective standing wave cavity
modes (as opposed to backscattering phase $\xi$) thus ensuring
azimuthal alignment of the QD with a field antinode of one of the
standing wave modes.  Note that in the example of Sec.
\ref{sec:QME_sims_1} (Fig. \ref{fig:QME_sim_results_1}) in which
$\beta>g_{0}$, it is the phase of $\beta$ ($\xi$) which determines
the position of the standing wave field antinodes with respect to
the QD.

\section{Intensity correlation function calculations}
\label{sec:g2_calcs}

Of additional interest is the behavior of the (normalized) intensity correlation function $g^{(2)}(\tau)$, whose value
can indicate nonclassical characteristics of the cavity field \cite{ref:Mandel_Wolf}, and is thus of essential
importance in the characterization of QD-cavity based devices such as single photon sources
\cite{ref:Michler,ref:Santori,ref:McKeever,ref:Birnbaum}.  Furthermore, intensity correlations of the cavity field (and
other higher-order correlations) are sensitive to the energy levels of multi-photon states of the system, and thus
provide further information about the system beyond the weak driving limit studied above.  This is particularly
important in the case of WGM cavities, in which the presence of a double-peaked spectrum typically associated with
Rabi-splitting cannot by itself be regarded as evidence for strong coupling.  Here we analyze the intensity correlations
of a coupled QD-cavity system, the cavity containing a pair of nearly-degenerate WGM modes as in the analysis of the
previous sections.

A general definition for any (stationary) two-time intensity correlation function in our system is \cite{ref:Walls_Milburn,ref:Gerry}:

\begin{center}
\begin{equation}
\label{eq:g2_gen_defn}
g^{(2)}_{a,b}({\tau})=\underset{t\rightarrow\infty}{lim}\frac{\langle\hat{a}^{\dagger}(t)\hat{b}^{\dagger}(t+\tau)\hat{b}(t+\tau)\hat{a}(t)\rangle}{\langle\hat{a}^{\dagger}(t)\hat{a}(t)\rangle\langle\hat{b}^{\dagger}(t+\tau)\hat{b}(t+\tau)\rangle}
\end{equation}
\end{center}

\noindent where $\hat{a}$ and $\hat{b}$ are the field annihilation
operators for modes $a$ and $b$, which can be the cavity traveling
wave modes (labeled $cw/ccw$) or standing wave modes (labeled
$sw_{1}/sw_{2}$). Here, it is assumed that steady-state has been
reached (i.e., $t\rightarrow\infty$), so $g^{(2)}_{a,b}(\tau)$ is
the stationary two-time correlation function, and is a function of
the time delay $\tau$ only. We calculate $g^{(2)}_{a,b}(\tau)$ by
applying the quantum regression theorem \cite{ref:Carmichael} and
numerically integrating the quantum master equation (eqns.
(\ref{eq:empty_cavity_Hamiltonian})-(\ref{eq:loss_terms_3}))
\cite{ref:Tan1,ref:Tan2}. In what follows, we initially focus on
calculating $g^{(2)}_{ccw,ccw}(\tau)$, the two-time intensity
auto-correlation function for the $\emph{counterclockwise}$ WGM
field operator, $\hat{a}_{\text{CCW}}$. Due to phase-matching, the
reflected signal from the cavity is proportional to
$\hat{a}_{\text{CCW}}$, allowing such intensity correlations to be
measured in practice.

We begin by considering the set of parameters studied in the
steady-state transmission and reflection spectrum of Fig.
\ref{fig:QME_sim_results_1}(c), where $\xi=0$ so that $\beta$ is
purely real and positive, and where $\Delta\omega_{ac}=-\beta$ so
that the QD is tuned to resonance with the empty cavity lower
frequency standing wave mode which it is spatially aligned with.
$g^{(2)}_{ccw,ccw}(\tau)$ is calculated in three instances, with
each case corresponding to a probe field frequency $\omega_{l}$
tuned onto resonance with one of the three resonance peaks in the
coupled cavity-QD reflection spectrum of Fig.
\ref{fig:QME_sim_results_1}(c).  The results are shown in Fig.
\ref{fig:QME_g2_tau_1}. For probe frequencies
$\omega_{l}=-\beta{\pm}g_{0}\sqrt{2}$ (Fig.
\ref{fig:QME_g2_tau_1}(a)-(b)), photon antibunching and
sub-Poissonian statistics are predicted. This antibunching is a
result of the anharmonicity of the Jaynes-Cummings system; once
the system absorbs a photon at $-\beta{\pm}g_{0}\sqrt{2}$,
absorption of a second photon at the same frequency is not
resonant with the higher excited state of the system
\cite{ref:Birnbaum}.  The degree of antibunching is a function of
the specific system parameters chosen, and $g^{(2)}_{ccw,ccw}(0)$
approaches zero more closely as $g_{0}$ further exceeds the rates
$\kappa_{T}$ and $\gamma_{\perp}$. In this case, the difference in
$g^{(2)}_{ccw,ccw}(\tau)$ for probe frequencies $\omega_{l}=-\beta
\pm g_{0}\sqrt{2}$ is a result of the asymmetry in the spectrum of
the system due to the presence of the nominally uncoupled
high-frequency standing wave mode (asymmetry in the probe
frequency detuning from the CCW traveling wave mode's natural
frequency also plays a role here, and has a persistent effect upon
$g^{(2)}_{ccw,ccw}(\tau)$ even for increasing mode-coupling
($\beta$, $g_{0}$) and decreasing dephasing ($\kappa_{T}$,
$\gamma_{\perp}$)).  For a probe frequency resonant with the third
reflection peak at $\omega_{l}=\beta$ (Fig.
\ref{fig:QME_g2_tau_1}(c)), $g^{(2)}_{ccw,ccw}(\tau)$ is
essentially unity for all times and the reflected light from the
cavity is nearly Poissonian due to the spatial misalignment, and
resulting de-coupling, of the QD from the high-frequency cavity
mode light field.

\begin{figure*}
\begin{center}
\epsfig{figure=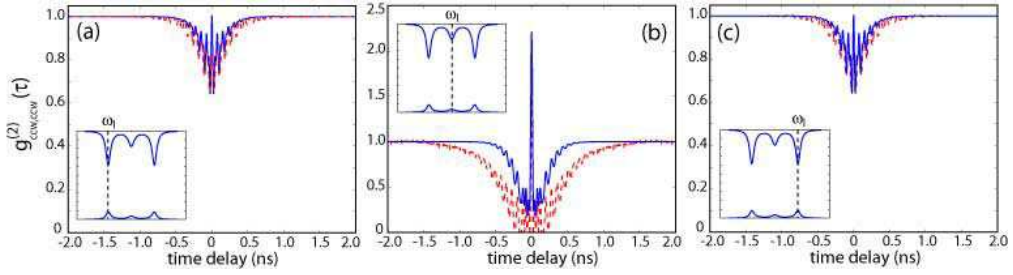, width=0.75\linewidth}
\caption{Normalized second order auto-correlation function (solid
blue line) $g^{(2)}_{ccw,ccw}(\tau)$ for the counterclockwise
propagating cavity mode for the parameters
  $\{g_{0},\beta,\kappa_{T},\kappa_{e},\gamma_{\parallel},\gamma_p\}/2\pi=\{6,i9.6,1.2,0.44,0.16,2.4\}$ GHz,
  $\Delta\omega_{ac}=0$, and whose transmission/reflection spectrum, originally shown in Fig. \ref{fig:QME_sim_results_3}(a), is
re-displayed here in the plot insets.
  (a) $\omega_{l}/2\pi=-12.8$ GHz, (b) $\omega_{l}/2\pi=0$ GHz, and (c) $\omega_{l}/2\pi=12.8$ GHz.  The additional red
  dashed line plots correspond to a QD with no non-radiative dephasing ($\gamma_{p}/2\pi=0$ GHz).}
\label{fig:QME_g2_tau_2}
\end{center}
\end{figure*}

\begin{figure*}
\begin{center}
\epsfig{figure=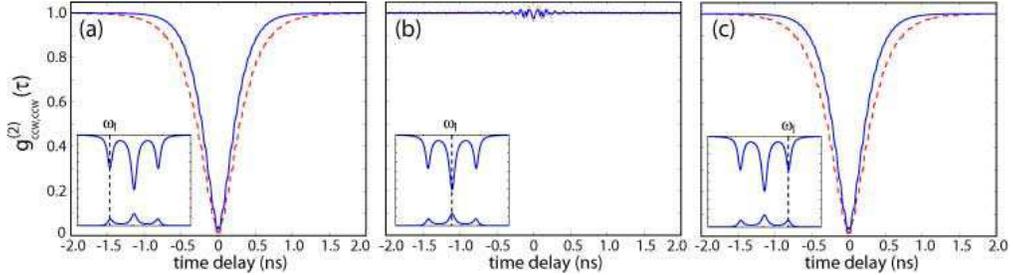, width=0.75\linewidth}
\caption{Normalized second order correlation function (solid blue
  line) $g^{(2)}_{ccw,ccw}(\tau)$ for the counterclockwise propagating cavity mode for the parameters
  $\{g_{0},\beta,\kappa_{T},\kappa_{e},\gamma_{\parallel},\gamma_p\}/2\pi=\{6,0,1.2,0.44,0.16,0.7\}$ GHz,
  $\Delta\omega_{ac}=0$, and whose transmission/reflection spectrum, originally shown in Fig. \ref{fig:QME_solns_no_splitting}, is re-displayed here in the plot insets.
  (a) $\omega_{l}=-g_{0}\sqrt{2}$, (b) $\omega_{l}=0$, and (c) $\omega_{l}=g_{0}\sqrt{2}$.  The additional red dashed
  line plots correspond to a QD with no non-radiative dephasing ($\gamma_{p}/2\pi=0$ GHz).} \label{fig:QME_g2_tau_3}
\end{center}
\end{figure*}

We next examine the parameter set explored in Fig.  \ref{fig:QME_sim_results_3}(a), where $\xi=\pi/2$, so that both
standing wave cavity modes are spatially coupled to the QD. In addition, $\Delta\omega_{ac}=0$, so that the modes are
equally and oppositely detuned from the QD.  Once again, we calculate $g^{(2)}_{ccw,ccw}(\tau)$ for three cases, with
each case corresponding to $\omega_{l}$ on resonance with one of the three peaks in the reflection spectrum of Fig.
\ref{fig:QME_sim_results_3}(a).  The results, shown in Fig.  \ref{fig:QME_g2_tau_2}, indicate mild antibunching for
$\omega_{l}/2\pi=\pm{12.8}$ GHz (the leftmost and rightmost peaks in the reflection spectrum) and for $\gamma_{p}$=0.
Higher levels of non-radiative dephasing lead to a washing out of the antibunching, though the field exhibits
sub-Poissonian statistics on a time scale of $\sim1/(\kappa_{T}+\gamma_{\perp})/2$. For $\omega_{l}$=0 (Fig.
\ref{fig:QME_g2_tau_2}(b)), the calculation predicts photon bunching.  This occurs because the resonance at zero
detuning only appears when at least one photon is in the cavity and is coupled to the QD, which then allows for an
additional photon at this frequency to be stored in the cavity and reflected.  The high-frequency oscillations in
$g^{(2)}_{ccw,ccw}(\tau)$, and resulting narrow super-Poissonian central peak about $g^{(2)}_{ccw,ccw}(0)$, are a result
of interference effects created by beating between the two cavity modes that are excited in this case (note that the
central resonance peak in the cavity-QD spectrum is predominantly atomic-like, and excitation through the optical
channel effectively excites the two detuned peaks, which are primarily photonic in nature).

\begin{figure*}[t]
\begin{center}
\epsfig{figure=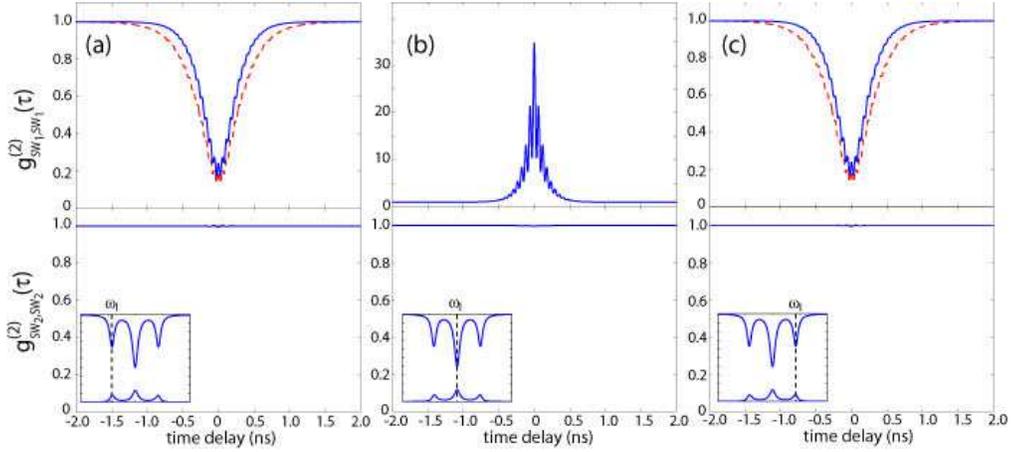, width=0.75\linewidth}
\caption{Normalized second order correlation functions (solid blue
  line) $g^{(2)}_{sw_{1},sw_{1}}(\tau)$ (top plots) and $g^{(2)}_{sw_{2},sw_{2}}(\tau)$ (bottom plots) for the parameters
  $\{g_{0},\beta,\kappa_{T},\kappa_{e},\gamma_{\parallel},\gamma_p\}/2\pi=\{6,0,1.2,0.44,0.16,0.7\}$ GHz,
  $\Delta\omega_{ac}=0$, and whose transmission/reflection spectrum,
  originally shown in Fig. \ref{fig:QME_solns_no_splitting}, is re-displayed here in the plot
  insets. (a) $\omega_{l}=-g_{0}\sqrt{2}$, (b) $\omega_{l}=0$, and
 (c) $\omega_{l}=g_{0}\sqrt{2}$.  The additional red dashed
  line plots correspond to a QD with no non-radiative dephasing ($\gamma_{p}/2\pi=0$ GHz).} \label{fig:QME_g2_cw_sw_tau_3}
\end{center}
\end{figure*}

Finally, we consider the parameter set explored in Fig.  \ref{fig:QME_solns_no_splitting}, where $\beta$=0 so that only the
QD couples the clockwise and counterclockwise modes together.  For the Rabi-split peaks centered at $\omega_{l}=\pm
g_{0}\sqrt{2}$, we see strong antibunching, as to be expected for a single QD coupled to a single cavity mode excited on
resonance with the Rabi-split peaks. At $\omega_{l}=0$, there are just minor oscillations about $g^{(2)}(\tau)$=1 due to
the weak and transient coupling of the resonant cavity mode with the QD. Comparison of this example with that of Fig.
\ref{fig:QME_g2_tau_2} illustrates well the added system information gained by studying intensity correlations of the
scattered light.  Although both systems look very similar when studying the amplitude of light transmission and
reflection intensity under weak-driving, the intensity correlations provide information about the spatial position of
the QD relative to each of the standing wave cavity modes and the relative strength of $g_{0}$ to $\beta$.

Up to this point we have considered only the two-time correlation
function for the counterclockwise propagating cavity mode.
Correlation functions for the clockwise propagating mode and
standing wave cavity modes can be determined through formulas
analogous to eqn. (\ref{eq:g2_gen_defn}), and can provide further
insight into the appropriateness of the standing wave mode
picture. Figure \ref{fig:QME_g2_cw_sw_tau_3} shows the results of
two-time intensity correlation calculations for the set of
parameters considered in Fig. \ref{fig:QME_g2_tau_3}, where now we
have plotted the intensity auto-correlation function for the
standing wave modes. The results are consistent with the standing
wave mode picture of atom-cavity interaction:  mode $sw_{1}$ is
spatially aligned with the QD, and hence
$g^{(2)}_{sw_{1},sw_{1}}(\tau)$ shows significant photon
antibunching at the Rabi-split frequencies $\omega_{l}=\pm
g_{0}\sqrt{2}$ and off-resonance bunching\cite{ref:Birnbaum} at
$\omega_{l}=0$ (top plots in Fig.
\ref{fig:QME_g2_cw_sw_tau_3}(a)-(c)), while standing wave mode
$sw_{2}$ is spatially misaligned from the QD and
$g^{(2)}_{sw_{2},sw_{2}}(\tau)$ is essentially unity for all drive
frequencies (bottom plots in Fig.
\ref{fig:QME_g2_cw_sw_tau_3}(a)-(c)).

In addition to the autocorrelation calculations presented thus
far, there are a number of other investigations of non-classical
behavior within this system that may be of interest. For example,
mixed-mode correlation functions can give insight into
entanglement between the two cavity modes and the potential for
generating non-classical states such as those employed in studies
of the Einstein-Podolsky-Rosen paradox
\cite{ref:Walls_Milburn,ref:Rauschenbeutel}. Squeezing, which has
been studied in the context of the Jaynes-Cummings system by a
number of authors
\cite{ref:Meystre,ref:Carmichael3,ref:Kuklinski,ref:Kimble,ref:Nha},
is also a potential topic for further study.  The strong coupling
of two cavity modes to a single QD, in the presence of
background-mediated intermodal coupling, may yield important
differences from previously studied systems.  Furthermore,
generating squeezed light or other non-classical fields in a
microchip-based geometry could be of technological importance.  In
the Appendix, we present some preliminary calculations on the
above topics which indicate the degree to which such non-classical
behavior may be exhibited in these devices. A more authoritative
treatment of these topics requires a systematic investigation of
different parameter regimes for
$\{g_{0},\beta,\kappa_{T},\kappa_{e},\gamma_{\parallel},\gamma_p\}$,
driving field strength and frequency, and excitation channel
(coupling to the cavity mode versus coupling to the QD directly),
and is beyond the scope of this paper.

\section{Summary}

We have extended the standard quantum master equation model for a two-level system coupled to the mode of an
electromagnetic cavity to better reflect the situation that occurs in realistic semiconductor microdisk cavities.  In
this model the quantum dot, still treated as a two-level system, is coupled to two cavity modes corresponding to
clockwise and counterclockwise propagating whispering-gallery modes of the disk.  These two modes are in turn passively
coupled to each other through surface roughness, characterized by a backscatter parameter $\beta$.  We examine the
steady state behavior of the system for differing regimes of $\beta$, the QD-cavity coupling rate $g_{0}$, the cavity
decay rate $\kappa_{T}$, and the quantum dot dephasing rate $\gamma_{\perp}$.  In particular, we consider conditions for
which standing wave cavity modes form, how the magnitude of the different system rates and the phase of $\beta$
determine the nodes and antinodes of the cavity modes with respect to the quantum dot, and the resulting QD-cavity
coupling.  It is anticipated that this analysis will be useful in the interpretation of experimental spectra from a
waveguide-coupled whispering-gallery-mode microcavity strongly coupled to a single two-level system such as the exciton
state of a self-assembled quantum dot.

\section{Acknowledgements}

K.S. acknowledges the Hertz Foundation and K.S. and O.P.
acknowledge the Caltech Center for the Physics of Information for
support of this work.

\appendix*
\section{Non-classical correlations and squeezing}
\label{app:squeezing}

\begin{figure*}[t]
\begin{center}
\epsfig{figure=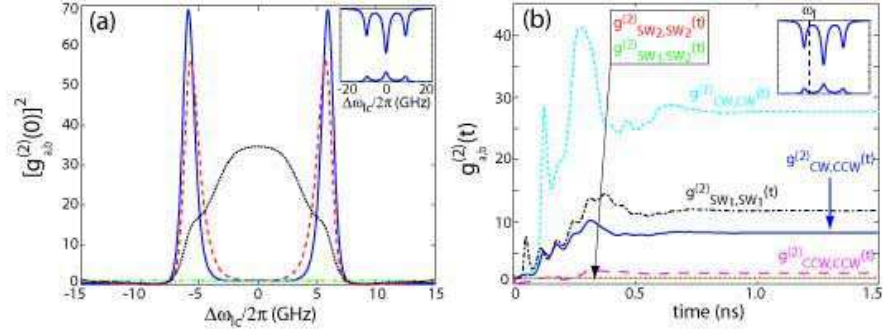, width=0.65\linewidth} \caption{Second
order correlation functions for the parameters
  $\{g_{0},\beta,\kappa_{T},\kappa_{e},\gamma_{\parallel},\gamma_p\}/2\pi=\{6,0,1.2,0.44,0.16,0.7\}$ GHz,
  $\Delta\omega_{ac}=0$, and whose transmission/reflection spectrum,
  originally shown in Fig. \ref{fig:QME_solns_no_splitting}, is re-displayed here in the plot
  insets.  (a) $(g^{(2)}_{cw,ccw}(0))^2$ (solid blue line), $g^{(2)}_{cw,cw}(0)\ast
  g^{(2)}_{ccw,ccw}(0)$ (dashed red line), $(g^{(2)}_{sw_{1},sw_{2}}(0))^2$ (dash-dotted green line), and $g^{(2)}_{sw_{1},sw_{1}}(0)\ast
  g^{(2)}_{sw_{2},sw_{2}}(0)$ (dotted black line) as a function of
  laser-cavity detuning. (b) $g^{(2)}_{cw,ccw}(t)$, $g^{(2)}_{cw,cw}(t)$, $g^{(2)}_{ccw,ccw}(t)$,
  $g^{(2)}_{sw_{1},sw_{2}}(t)$, $g^{(2)}_{sw_{1},sw_{1}}(t)$, and $g^{(2)}_{sw_{2},sw_{2}}(t)$ for excitation at $\omega_{l}/2\pi=-6$
  GHz, and when both cavity modes are initially in
  the vacuum state and the QD is in its ground state.} \label{fig:QME_g2_mixed_mode}
\end{center}
\end{figure*}

\subsection{Mixed-mode correlation function}

We reconsider the parameter set of Fig. \ref{fig:QME_g2_tau_3}, where $\beta$=0, so that only the QD is coupling the two
cavity modes, and calculate the mixed-mode correlation functions $g^{(2)}_{cw,ccw}(\tau)$ and $g^{(2)}_{sw_{1},sw_{2}}(\tau)$
for the same set of parameters.  We focus on solutions for $\tau$=0, which can be obtained entirely from the
steady-state density matrix, and examine the behavior of $g^{(2)}_{a,b}(0)$ as a function of driving frequency. The
Cauchy-Schwarz inequality

\begin{center}
\begin{equation}
\label{eq:g2_Cauchy_Schwarz} \Bigl(g^{(2)}_{a,b}\Bigr)^2\leq
g^{(2)}_{a,a}\ast g^{(2)}_{b,b}
\end{equation}
\end{center}

\noindent is violated when non-classical correlations exist
between the two modes $a$ and $b$
\cite{ref:Walls_Milburn,ref:Gerry}.  As we see in Fig.
\ref{fig:QME_g2_mixed_mode}(a), this inequality is violated for
the traveling wave modes at particular choices of $\omega_{l}$, so
that non-classical correlations between the two modes can occur in
this system. On the other hand, for the standing wave modes, no
quantum correlations exist, as mode $sw_{2}$ is not coupled to the
QD, so that
$g^{(2)}_{sw_{1},sw_{2}}(0)=g^{(2)}_{sw_{2},sw_{2}}(0)=1$ for all
$\omega_{l}$. The transient (i.e., non-steady-state) behavior of
the mixed and single mode correlation functions are shown in Fig.
\ref{fig:QME_g2_mixed_mode}(b).  Here we plot
$g^{(2)}_{cw,ccw}(t)$, $g^{(2)}_{sw_{1},sw_{2}}(t)$,
$g^{(2)}_{cw,cw}(t)$, $g^{(2)}_{ccw,ccw}(t)$,
$g^{(2)}_{sw_{1},sw_{1}}(t)$, and $g^{(2)}_{sw_{2},sw_{2}}(t)$ for
$\omega_{l}/2\pi=-6$ GHz (where the Cauchy-Schwarz inequality is
nearly maximally violated), with the $t=0$ initial state
consisting of both cavity modes in the vacuum state and the QD in
its ground state.  The calculations indicate that steady-state
behavior is achieved after $\sim$1 ns, corresponding to the
system's average decay time
($\sim1/(\kappa_{T}+\gamma_{\perp})/2$), with violations of the
Cauchy-Schwarz inequality for the traveling wave modes occurring
after only $\sim$0.2 ns.

\subsection{Squeezing}

\begin{figure*}
\begin{center}
\epsfig{figure=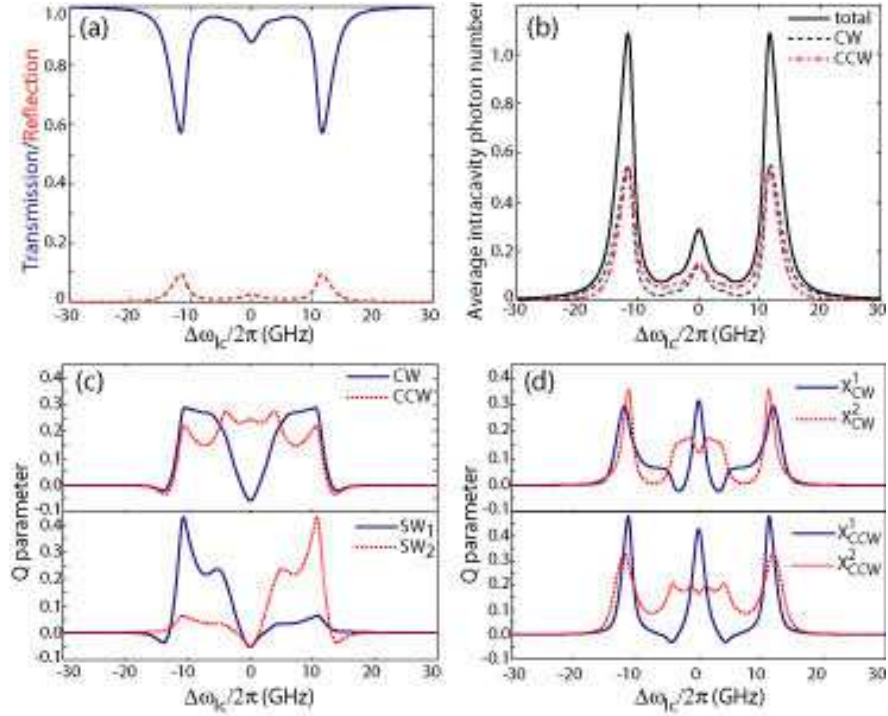, width=0.65\linewidth}
\caption{Non-classical properties of the microdisk cavity-QD
system under more intense driving ($\sim$30 photons/ns drive
power), and with system parameters
$\{g_{0},\beta,\kappa_{T},\kappa_{e},\gamma_{\parallel},\gamma_p\}/2\pi=\{6,i9.6,1.2,0.44,0.16,0.7\}$
GHz. (a) Cavity transmisson/reflection as a function of driving
field frequency $\omega_{l}$, (b) total (solid black line), CW
mode (dashed blue line), and CCW mode (dot-dashed red line)
intracavity photon number as a function of $\omega_{l}$,
 (c) Q parameter for the intracavity photon number in the (top)
  traveling wave WGMs and (bottom) standing wave WGMs. (d) Q parameter for the $X^{1}$ and $X^{2}$ quadratures of the (top)
  cw and (bottom) ccw traveling wave mode.} \label{fig:QME_Q_param}
\end{center}
\end{figure*}

As has been observed by several other authors in studies of single
mode cavity QED \cite{ref:Meystre,ref:Kuklinski}, squeezing in the
field quadratures can occur, however, with the amount of squeezing
typically small ($<20\%$) unless large intracavity photon numbers
($>10$) are achieved \footnote{It should be noted that Refs.
\cite{ref:Meystre,ref:Kuklinski} study the
  Jaynes-Cummings system in absence of a driving field.  Strong driving fields significantly affect the structure of the
  Jaynes-Cummings system, causing drive-strength-dependent Stark shifts of the system eigenergies \cite{ref:Alsing2}. We
  thus might expect strong driving fields to have an appreciable affect on squeezing and photon statistics.  In
  particular, Alsing and co-workers \cite{ref:Alsing2} have shown that squeezed, displaced number states are
  eigenstates of the total Hamiltonian for the driven Jaynes-Cummings model.}. The basic reason for this is that the
nonlinear interaction that generates squeezing in the
Jaynes-Cummings system is that of the electromagnetic field
coupling to a saturable oscillator (the QD); this implies that the
intracavity field has to be strong enough for QD saturation
effects to be appreciable. For our system, situations where only
one of the standing wave modes is coupled to the QD (as in Figs.
\ref{fig:QME_g2_tau_1} and \ref{fig:QME_g2_tau_3}, for example)
essentially reduce to that of the single mode cavity QED case, and
we expect qualitatively similar behavior.  A perhaps more
interesting example to study is that of Fig.
\ref{fig:QME_g2_tau_2}, where both standing wave modes are equally
coupled to a QD.  To achieve a reasonable intracavity photon
number, we increase the input driving field by approximately three
orders of magnitude over the weak drive fields we have used up to
this point, to a level of $\sim30$ photons/ns, so that the average
intracavity photon number (Fig. \ref{fig:QME_Q_param}(b)) peaks at
a value of $\sim$1. This results in the transmission and
reflection spectra shown at the top of Fig.
\ref{fig:QME_Q_param}(a).  In comparison to the
transmission/reflection spectra calculated in the weak driving
limit in Fig. \ref{fig:QME_sim_results_3}(a), we now begin to see
asymmetries in the transmission dips (reflection peaks) that are
associated with multi-photon transitions to excited states in the
Jaynes-Cummings spectrum and QD saturation effects.  In this
calculation, we are unable to numerically study higher drive
strengths due to the resulting large system size for these
two-mode cavities. To access higher driving fields using the same
computational resources, adopting a wavefunction-based approach
(i.e., the quantum Monte Carlo method) is one possibility.

We next consider fluctuations in the steady-state, internal cavity field
(squeezing in the external field, which can be investigated through the spectrum
of squeezing \cite{ref:Carmichael4}, for example, are not considered here but may be of future
interest).
First, we look at fluctuations in the photon number in mode $i$ by
calculating the Mandel Q parameter \cite{ref:Mandel_Wolf}

\begin{center}
\begin{equation}
\label{eq:Mandel_Q_1}
Q(\hat{n}_{i})=\frac{\text{Var}(\hat{a}_{i}^{\dagger}\hat{a}_{i})-\langle\hat{a}_{i}^{\dagger}\hat{a}_{i}\rangle}{\langle\hat{a}_{i}^{\dagger}\hat{a}_{i}\rangle}
\end{equation}
\end{center}

\noindent where for some operator $\hat{O}$,
$\text{Var}(\hat{O})=\langle\hat{O}^2\rangle-\langle\hat{O}\rangle^2$.
Figure \ref{fig:QME_Q_param}(c) shows the calculated Q parameter
as a function of driving field frequency for the $cw/ccw$
traveling wave modes (top) and $sw_{1}/sw_{2}$ standing wave modes
(bottom). These plots show $Q<0$ for certain driving frequencies,
indicating that sub-Poissonian photon number statistics can be
achieved, though the level of non-classicality is small ($\sim5
\%$). A calculation of $Q(\hat{n}_{\text{CW}})(t)$ for
$\omega_{l}$=0 ( with both cavity modes initially in the vacuum
state and the QD in its ground state) indicates that slightly
higher levels of non-classicality ($Q\sim -0.1$) can be achieved
before steady-state is reached.  Additional preliminary
calculations using a quantum Monte Carlo method to access higher
drive strengths have been performed, and show that $Q$ can
continue to decrease for larger driving fields.  For drive
strengths of $\sim300$ photons/ns (corresponding to an average
total intracavity photon number $\sim$1 at $\omega_{l}$=0),
$Q(\hat{n}_{\text{CW}})$ can reach $-0.35$ in its transient
(non-steady-state) behavior.

Similarly, one can examine fluctuations in the field quadratures.  For mode $i$, we define the quadrature operators
$\hat{X}_{i}^{1,2}$ by:

\begin{equation}
\begin{split}
\label{eq:quadrature_operators}
\hat{X}_{i}^{1}&=\frac{1}{2}(\hat{a}_{i}+\hat{a}_{i}^{\dagger}),\\
\hat{X}_{i}^{2}&=\frac{-i}{2}(\hat{a}_{i}-\hat{a}_{i}^{\dagger})
\end{split}
\end{equation}

\noindent The corresponding $Q$ parameter for quadrature $j$ of
mode $i$ is then \cite{ref:Gerry}:

\begin{center}
\begin{equation}
\label{eq:Mandel_Q_2}
Q_{i}^{j}=\frac{\text{Var}(\hat{X}_{i}^{j})-0.25}{0.25}
\end{equation}
\end{center}

\noindent From Fig. \ref{fig:QME_Q_param}(d), we see that small
amounts of quadrature squeezing in the $cw/ccw$ modes are apparent
for the conditions considered.

\bibliography{./PBG_1_25_2007}

\end{document}